\documentclass[aps,prb,twocolumn,floatfix,superscriptaddress]{revtex4-1}
\usepackage{graphicx,txfonts,bm}
\usepackage[colorlinks,citecolor=magenta,linkcolor=blue]{hyperref}
\begin{document}


\title{Exploring the exotic $f$ states of prototype compounds CeSb and USb}


\author{Haiyan Lu}
\email{luhaiyan@itp.ac.cn}
\affiliation{Science and Technology on Surface Physics and Chemistry Laboratory, P.O. Box 9-35, Jiangyou 621908, China}

\author{Qin Liu}
\affiliation{Science and Technology on Surface Physics and Chemistry Laboratory, P.O. Box 9-35, Jiangyou 621908, China}

\date{\today}

\begin{abstract}
To unravel the interplay between the strong electronic correlation and itinerant-localized dual nature in atypical $f$ electron systems, we employed the density functional theory in combination with the single-site dynamical mean-field theory to systematically investigate the electronic structures of CeSb and USb. We find that the 4$f$ states in CeSb are mostly localized which show a weak quasi-particle resonance peak near the Fermi level. Conversely, the 5$f$ electrons in USb display partially itinerant feature, accompanied by mixed-valence behavior and prominent valence state fluctuations. Particularly, the 4$f$ electronic correlations in CeSb are distinctly orbital-selective with strikingly renormalized $4f_{5/2}$ states, according to the low-energy behaviors of 4$f$ self-energy functions. It is believed that the strong electronic correlation and fantastic bonding of $f$ states contribute to elucidate the fascinating magnetism.
\end{abstract}


\maketitle

\section{Introduction\label{sec:introduction}}
Lanthanides and actinides with partially filled 4$f$ and 5$f$ shells exhibit abundant physical behaviors including heavy-fermion behavior~\cite{RevModPhys.56.755}, quantum criticality~\cite{Sachdev2011Quantum}, magnetic ordering~\cite{H1986Anisotropic,Lander1976Neutron,Lander1980Neutron,PhysRevLett.42.260}, unconventional superconductivity~\cite{Sarrao2002,Curro2005PuCoGa5} and mixed-valence states~\cite{RevModPhys.81.235}. It is generally believed that the properties are closely associated with electronic structure which arises from strong electronic correlation, large spin-orbital coupling and intricate crystal field splitting. Usually, the 4$f$ electrons of the rare earths are considered to be localized and unlikely to involve in bonding. On the other hand, the early actinides (from Ac to Np) with successive filled 5$f$ shell incline to be itinerant. Consequently, 5$f$ electrons tend to locate near the Fermi level and strongly participate in bonding, generating a plethora of interesting and unique physics~\cite{RevModPhys.81.235}.

\begin{figure}[ht]
\centering
\includegraphics[width=\columnwidth]{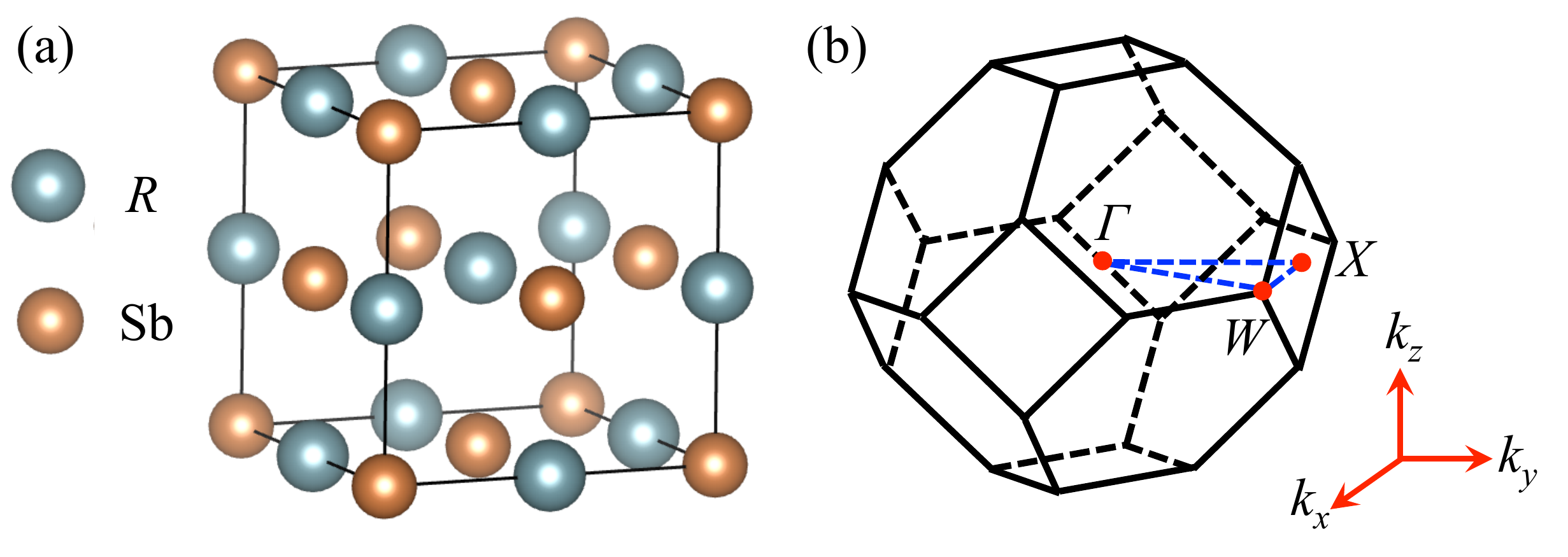}
\caption{(Color online). (a) Crystal structure of $R$Sb ($R$=Ce, U). (b) The first Brillouin zone for $R$Sb ($R$=Ce, U), where high-symmetry $k$ points are marked.
\label{fig:tstruct}}
\end{figure}

Special interests have been drawn to cerium antimonide (CeSb) and uranium antimonide (USb), which stabilize in the rock-salt (NaCl) type of crystal structure (see Fig.~\ref{fig:tstruct}) with lattice constant 6.388 {\AA}~\cite{Bartholin1977} and 6.191 {\AA}~\cite{Lander1976Neutron}, respectively. It is reported that CeSb is semimetal with a small overlapping of Ce-5$d$ and Sb-5$p$ states, and 4$f$ states are mostly localized far away from the Fermi level. It orders antiferromagnetically below $T_N$=16 K and undergoes further phase transitions with decreasing temperature. Meanwhile, USb develops antiferromagnetic phase just below $T_N$=213 K~\cite{Kuznietz1969,PhysRevLett.42.260,Rossat1980Magnetic}. It is guessed that the extraordinary properties mainly stem from Ce-4$f$ and U-5$f$ electrons.

The magnetic phases of CeSb have been extensively studied using transport, thermodynamic, neutron scattering and spectroscopic experiments~\cite{Hulliger1975Low,Fischer1978Magnetic,Meier1978Magnetic,Furrer1979Magnetic,Lebech1981Neutron,Rossat1985Neutron,H1986Anisotropic,Chattopadhyay1994High,Vogt1995The,Kohgi2000Physics} for half a century. In the cubic crystal field, 4$f_{5/2}$ state of Ce atom splits into doublet $\Gamma_7$ and quartet $\Gamma_8$ states with a small energy gap around 19 K$\sim$ 26 K~\cite{Busch1971Crystal}, which enables the random distribution of spins and the formation of complicated antiferromagnetic ordering. Moreover, the photoemission spectroscopy (PES), angle-resolved photoemission spectroscopy (ARPES), optical conductivity and dHvA quantum oscillation shed light on the evolution of the electronic structure through paramagnetic to antiferromagnetic phase transition~\cite{PhysRevB.56.13654,Takahashi199865,Ito2004Para,Takayama2011Magnetic}. On the theoretical side, the widely used $p-f$ mixing model~\cite{Takahashi2001High} has been utilized to elucidate the effect of electronic structure on magnetism. This result is based on the assumption that the cubic symmetry is preserved during reducing temperature from paramagnetic phase to antiferromagnetic phase. However, CeSb is confirmed to distort from cubic structure to tetragonal one at low temperature~\cite{Duan2007Electronic}. Hence a reliable model for this issue should consider both the temperature-dependent crystal structures and the corresponding crystal field splitting. Meanwhile, the large spin-orbital coupling and intricate crystal field splitting make the theoretical calculations more difficult. So far, first-principle studies supposing the totally localized 4$f$ states are unable to describe the subtle electronic structure of CeSb~\cite{JPSJ.76.044707,Sakai2007,PhysRevB.86.115116,Sakai2012}. 
 
USb possesses a relatively small specific heat coefficient~\cite{Rudigier1983Low,PhysRevB.32.4584}, remarkably large electrical resistivity~\cite{PhysRevB.30.6578} and large magnetic moment~\cite{Lander1995On}, indicative of the localized degree of 5$f$ states. The antiferromagnetic phase develops below 213 K with magnetic moments in (001) planes stacked in sequence +-+-~\cite{Kuznietz1969,PhysRevLett.42.260}. A second phase transition appears below 142 K with triple-${\bf k}$ magnetic order~\cite{Lander1995Possible}. The magnetic phase diagram tuned by temperature and pressure, magnetic phase transition and spin dynamics have been intensively explored~\cite{PhysRev.173.562,Kuznietz1969,PhysRevLett.42.260,Rossat1980Magnetic,Lander1980Neutron,H1985Magnetic,Osborn1988High,
Lander1995On,Lander1995Possible,Magnetic1996pressure,Braithwaite1998Pressure,Nuttall2002mag,PhysRevB.87.064421}. 
Theoretically, the layered Ising model~\cite{Rossat1980Magnetic} and exchange model~\cite{Monachesi1983Low} have been used to explicate the multi-${\bf k}$ antiferromagnetic structures. However, the somewhat simplified model can not be generalized to treat other magnetic phases. Besides, the traditional $p-f$ mixing model based on $c-f$ hybridization fails to adequately seize the essence of the magnetic phase~\cite{Kumigashira1999High,PhysRevB.61.15707,Takahashi2001High}, which reveals a probable novel mechanism.
Furthermore, previous relativistic spin-density-functional theory~\cite{PhysRevB.61.6246} has been adopted to survey the band structure and magnetism of USb. They not only calculated the spin and orbital moment, but also found the discrepancy between the computed Fermi surface and dHvA experimental data. It is supposed that the difference comes from the underestimation of the strong electronic correlation among 5$f$ electrons and the hypothesis of the completely localized 5$f$ electrons.
On the other hand, the electronic structures have been widely investigated through PES, ARPES and dHvA quantum oscillation experiments~\cite{BAPTIST198063,ERBUDAK1980134,Tang1992Resonant,Takahashi199865,Kumigashira1999High,PhysRevB.61.15707,Takahashi2001High,Kumigashira2000High,Suzuki1995,JPSJ.66.2764}. These results reveal the itinerant-localized character of 5$f$ states. For example, the detected 5$f^2$ atomic multiplets imply the localized behavior of 5$f$ electrons~\cite{PhysRevB.61.15707}. The optical measurement manifests the existence of a wide U-6$d$ and U-5$f$ hybridization band around the Fermi level, demonstrating the itinerant 5$f$ states~\cite{Schoenes1981}. Thus it is hard to depict the partially itinerant 5$f$ electrons regarding the oversimplified assumption of the completely localized or itinerant 5$f$ states. Previous electronic structure calculations within local-density approximation~\cite{PhysRevB.22.645,JPSJ.64.2294,JPSJ.65.3394} are inadequate to capture the strong correlation among the 5$f$ electrons. Till now, the credible theoretical investigation of electronic structure is still lacking.
 
The present paper aims to address the following issues. First of all, the localized 4$f$ electrons and itinerant-localized 5$f$ electrons remain long-standing issues. Secondly, the impact of strong correlation among $f$ electrons on electronic structure is not yet fully understood. Thirdly, the underlying physics behind the magnetic phase transition is still unclear. To answer the questions above, it is crucial to examine the detailed electronic structures of CeSb and USb to explore the itinerant-localized dual nature and electronic correlation of $f$ states. It is believed that these results will enrich our understanding about the $f$ electron systems and serve for further studies.

The rest of this paper is arranged as follows. In Sec.~\ref{sec:method}, the computational details of DFT + DMFT approach are introduced. In Sec.~\ref{sec:results}, the electronic band structures, total and partial $f$ density of states, hybridization functions, valence state fluctuations and $f$ self-energy functions are presented. In Sec.~\ref{sec:dis}, similarities and differences between CeSb and USb are discussed. Moreover, the possible relation between electronic structure and magnetism is addressed. Finally, Sec.~\ref{sec:summary} gives a brief summary.

\section{Methods\label{sec:method}}
The strong correlation of Ce-4$f$ states and U-5$f$ states should be taken into account to accurately describe the electronic structure of CeSb and USb. It is established that the traditional density functional theory (DFT) combined with dynamical mean-field theory (DMFT) is a non-perturbative many-body approach to treat the local interactions between electrons~\cite{RevModPhys.68.13}. This method has been successfully utilized to study many lanthanides and actinides materials~\cite{Goremychkin186,Haule2009Arrested}. In the present paper, we employ the DFT + DMFT method to carry out charge fully self-consistent calculations to examine the electronic structure in detail.

Generally, the DFT+DMFT approach maps the lattice model to a quantum impurity model self-consistently and solves the obtained quantum impurity model by using various quantum impurity solvers. The calculation is divided into the DFT and DMFT parts. The DFT calculation is conducted by using the \texttt{WIEN2k}~\cite{wien2k} code which implements a full-potential linear augmented plane wave (FP-LAPW) formalism. The Perdew-Burke-Ernzerhof (PBE) functional~\cite{PhysRevLett.77.3865} is chosen to express the exchange-correlation potential. The $k$-points mesh was $21 \times 21\times 21$. Besides, $R_{\text{MT}}K_{\text{MAX}} = 7.0$. In addition, the spin-orbital coupling was explicitly included. The convergence criteria for charge and energy reach $10^{-4}$ e and $10^{-4}$ Ry, respectively. The experimental crystal structures for CeSb~\cite{Bartholin1977} and USb~\cite{Lander1976Neutron} were used. Since the inverse temperature $\beta = 40$ ($T \sim 298.0$~K), it was reasonable to retain only the paramagnetic solutions. 

We employed the \texttt{EDMFTF} code~\cite{PhysRevB.81.195107}, which implements the DFT + DMFT computational engine and the corresponding quantum impurity solvers, to study the obtained DFT + DMFT Hamiltonian. The constructed multi-orbital quantum impurity models were solved using the hybridization expansion continuous-time quantum Monte Carlo impurity solver (dubbed as CT-HYB)~\cite{RevModPhys.83.349,PhysRevLett.97.076405}. The Coulomb interaction strength $U$ and the Hund's exchange parameter $J$ are 6.0 eV and 0.7 eV, respectively. In order to simplify the calculations, we not only utilized the good quantum numbers $N$ and $J_z$ to reduce the sizes of matrix blocks of the local Hamiltonian, but also made a truncation for the local Hilbert space. It is emphasized that the atomic eigenstates were set with $N \in$ [0, 3] for Ce and with $N \in$ [0, 4] for U. Lastly, the lazy trace evaluation trick was applied to accelerate the Monte Carlo sampling further.

We performed charge fully self-consistent DFT + DMFT calculations, i.e., the correlation-corrected density matrix $\rho$ was built in the DMFT part, and then fed back to the DFT part to generate a new Kohn-Sham Hamiltonian $\hat{H}_{\text{KS}}$. Of the order of 60 DFT + DMFT iterations were required to obtain good convergence for the chemical potential $\mu$, charge density $\rho$, and total energy $E_{\text{DFT + DMFT}}$. The Matsubara self-energy functions $\Sigma(i\omega_n)$ generated in the last 10 DFT + DMFT iterations were collected and stored for further postprocessing.

\section{Results\label{sec:results}}

\begin{figure}[th]
\centering
\includegraphics[width=0.5\textwidth]{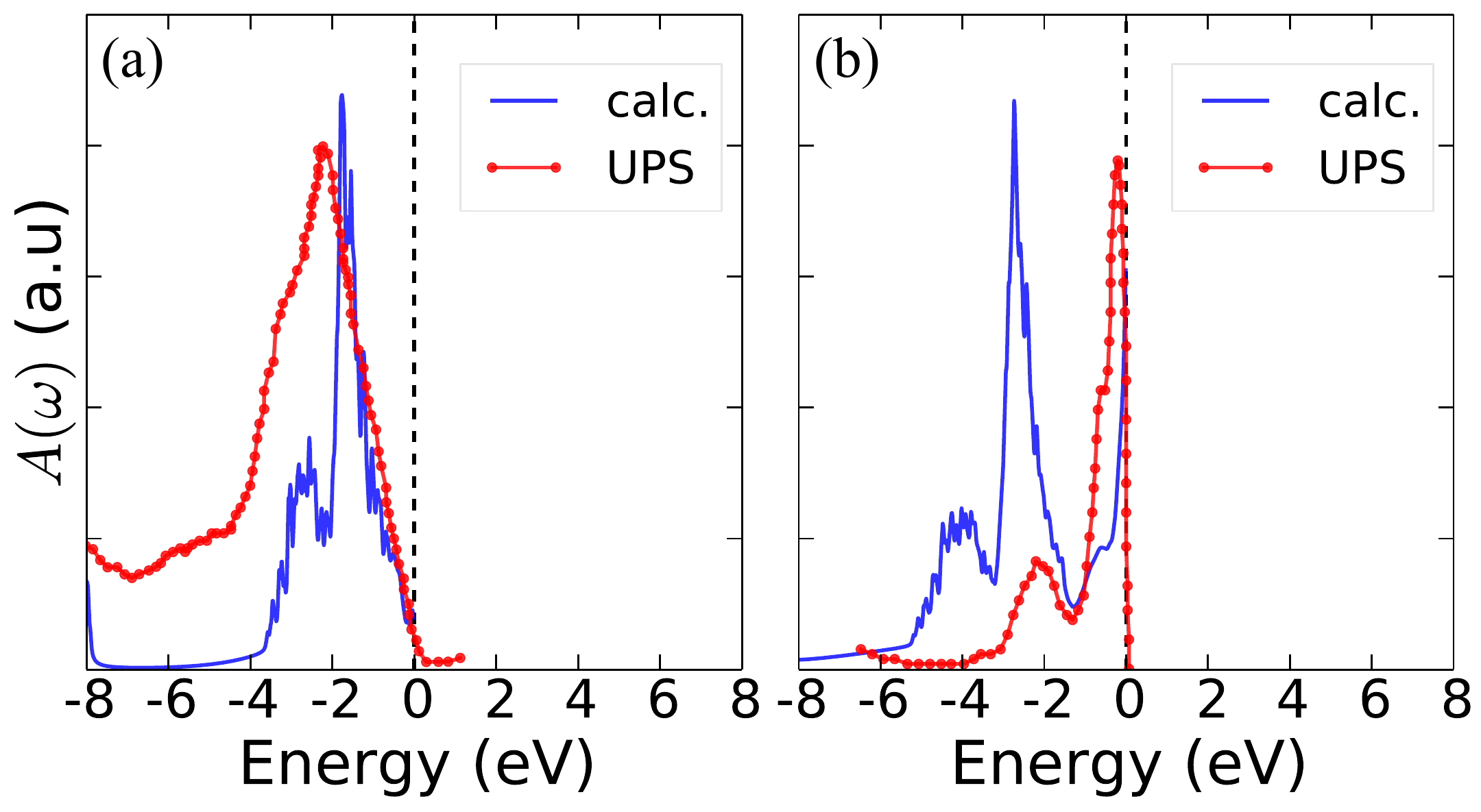}
\caption{(Color online). Comparisons of theoretical and experimental density of states for CeSb (a) and USb (b), respectively. In panel (a), the UPS data (filled red circles) are taken from Ref.~[\onlinecite{PhysRevB.18.4433}]. In panel (b), the experimental UPS data are taken from Ref.~[\onlinecite{Reihl1982Electronic,Reihl1987Photoelectron}]. The Fermi levels $E_{\text{F}}$ are represented by vertical dashed lines. Notice that the spectral data have been rescaled and normalized for a better visualization. \label{fig:tdos_exp}}
\end{figure}

\begin{figure*}[ht]
\centering
\includegraphics[width=\textwidth]{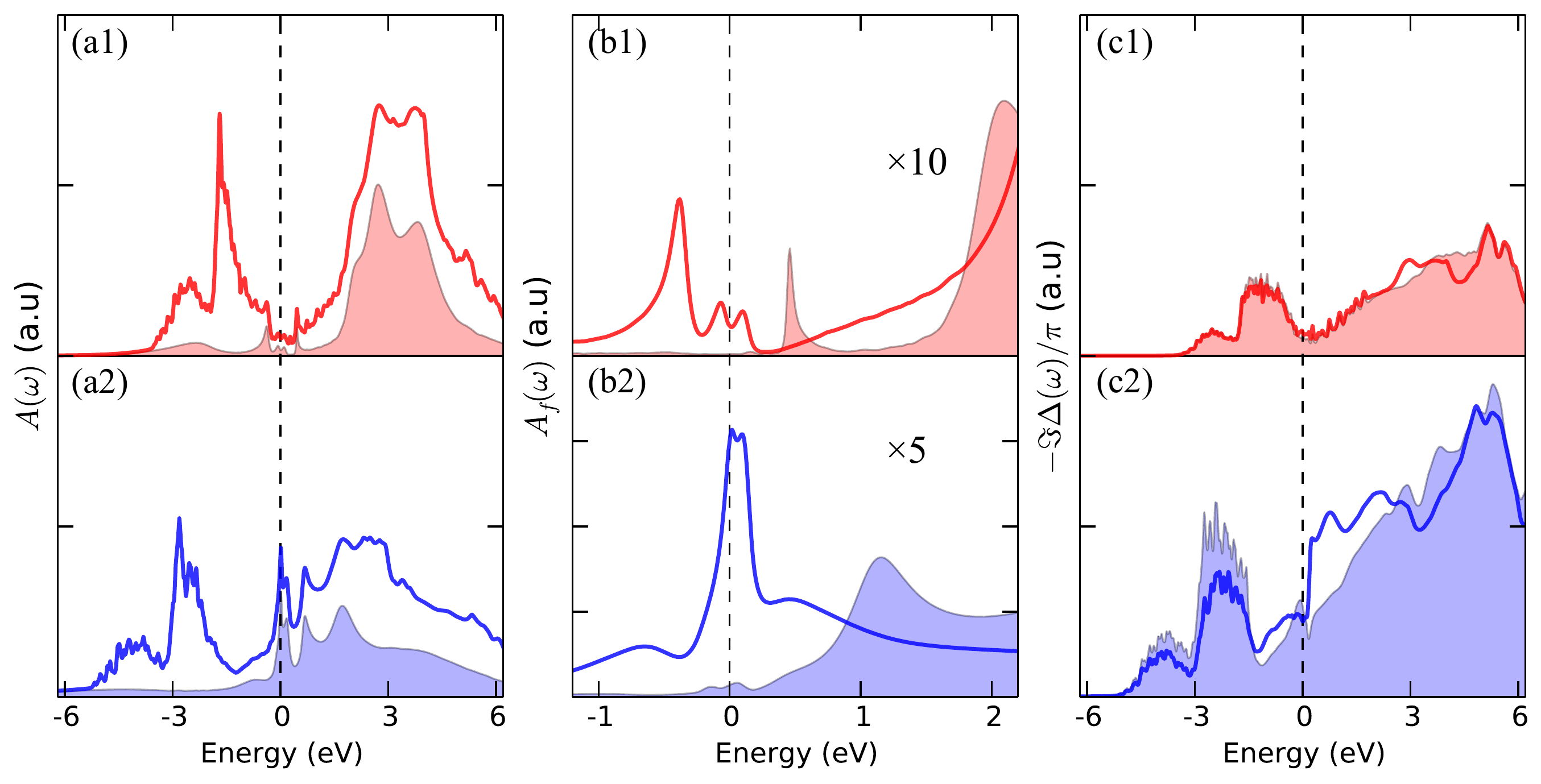}
\caption{(Color online). Electronic density of states of CeSb and USb obtained by DFT + DMFT calculations. Total density of states (thick solid lines) and partial $f$ density of states (color-filled regions) for CeSb (a1) and USb (a2), respectively. $f_{5/2}$ (thick solid lines) and $f_{7/2}$ (color-filled regions) for CeSb (b1) and USb (b2), respectively. The data presented in this figure are rescaled for a better view. Hybridization functions of $f_{5/2}$ (thick solid lines) and $f_{7/2}$ states (color-filled regions) for CeSb (c1) and USb (c2), respectively. The data presented in this figure are also rescaled. The vertical dashed lines denote the Fermi level. \label{fig:tdos}}
\end{figure*}

\begin{figure}[ht]
\centering
\includegraphics[width=0.5\textwidth]{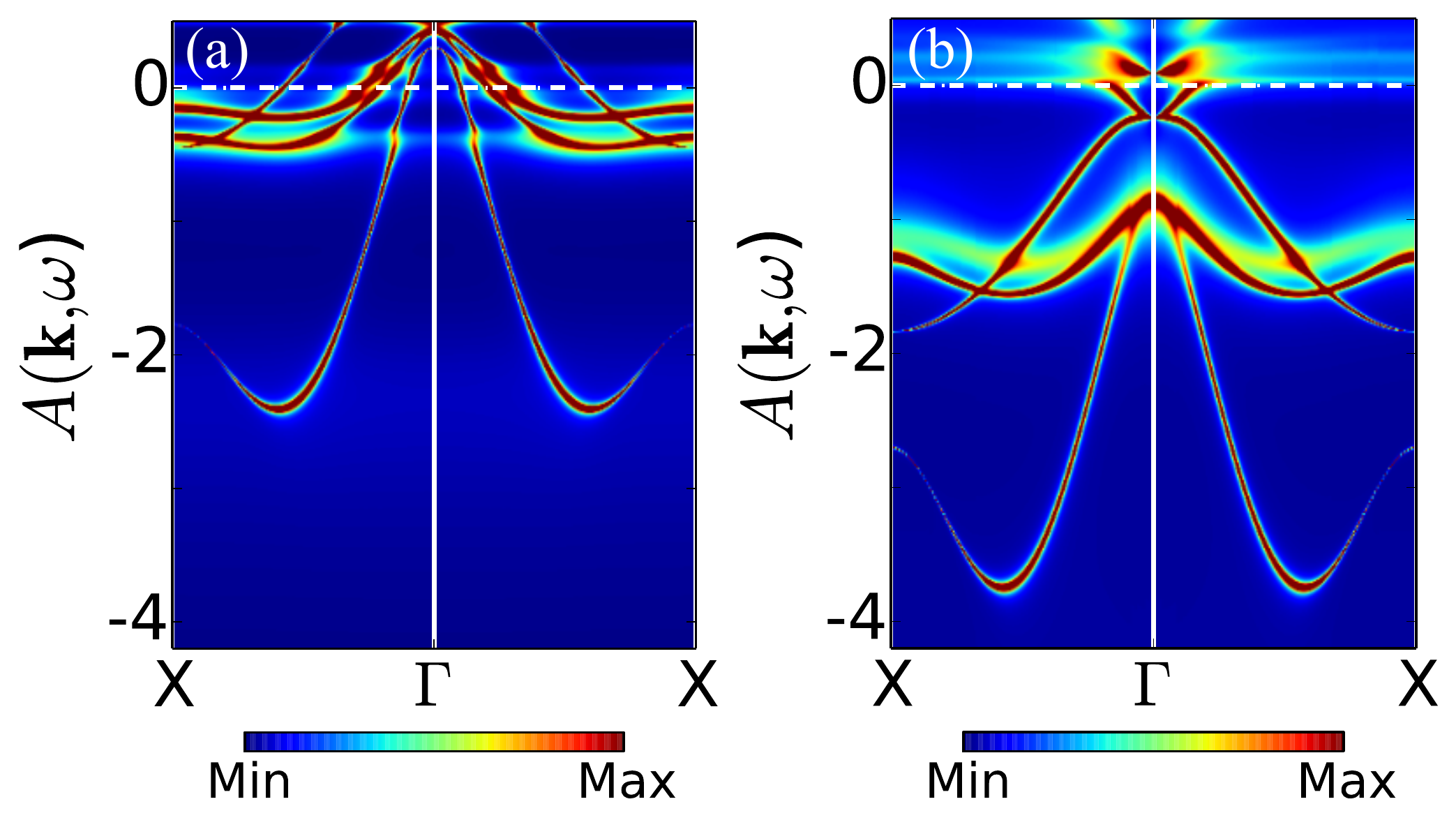}
\caption{(Color online). Momentum-resolved spectral functions $A(\mathbf{k},\omega)$ of CeSb (a) and USb (b) under ambient pressure. The horizontal lines denote the Fermi level. 
\label{fig:akw}}
\end{figure}

\begin{figure}[ht]
\centering
\includegraphics[width=0.5\textwidth]{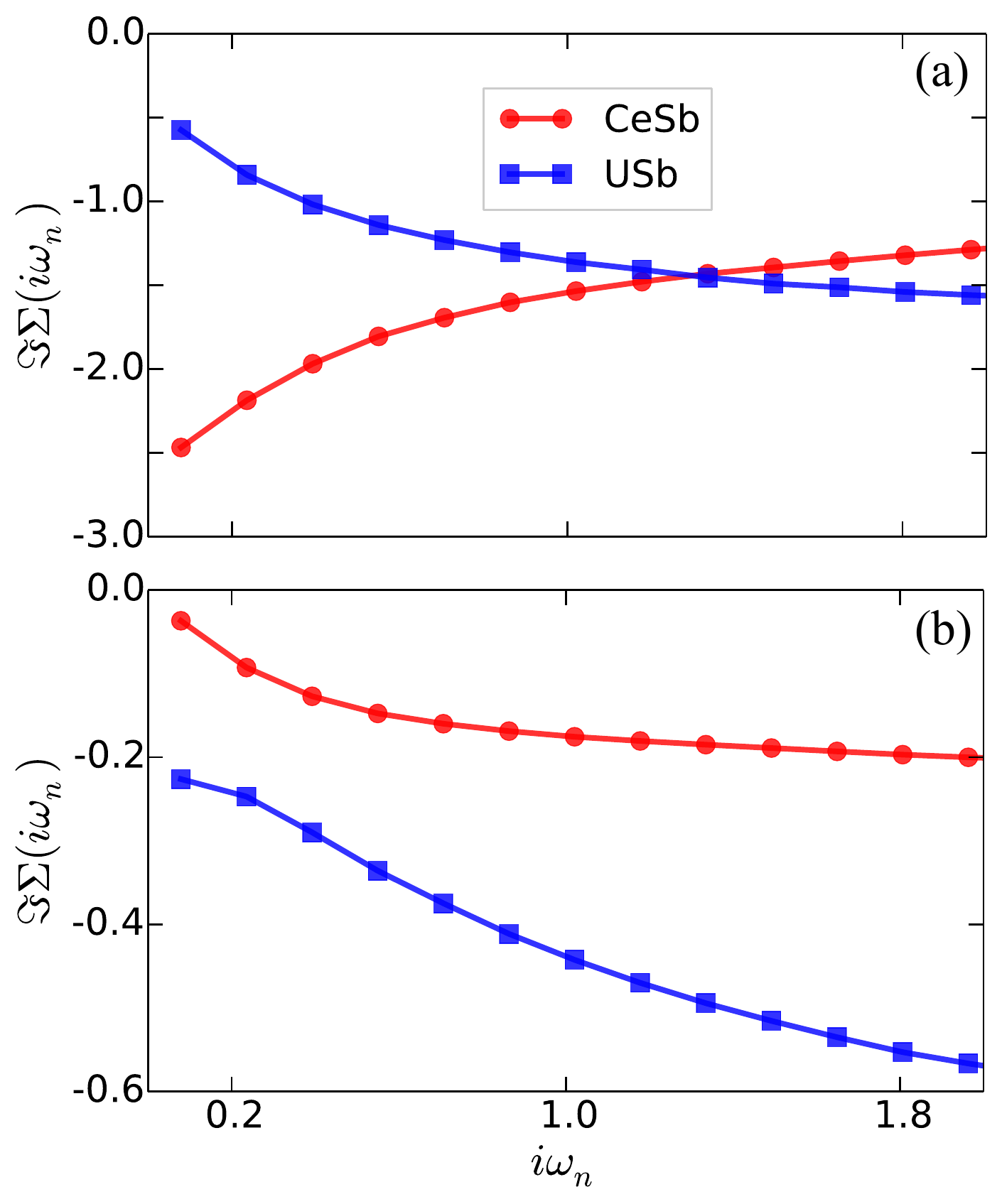}
\caption{(Color online). Imaginary parts of Matsubara self-energy functions of CeSb and USb derived by DFT + DMFT calculations. (a) $f_{5/2}$ components. (b) $f_{7/2}$ components. \label{fig:tsigma}}      
\end{figure}

\begin{figure}[ht]
\centering
\includegraphics[width=0.5\textwidth]{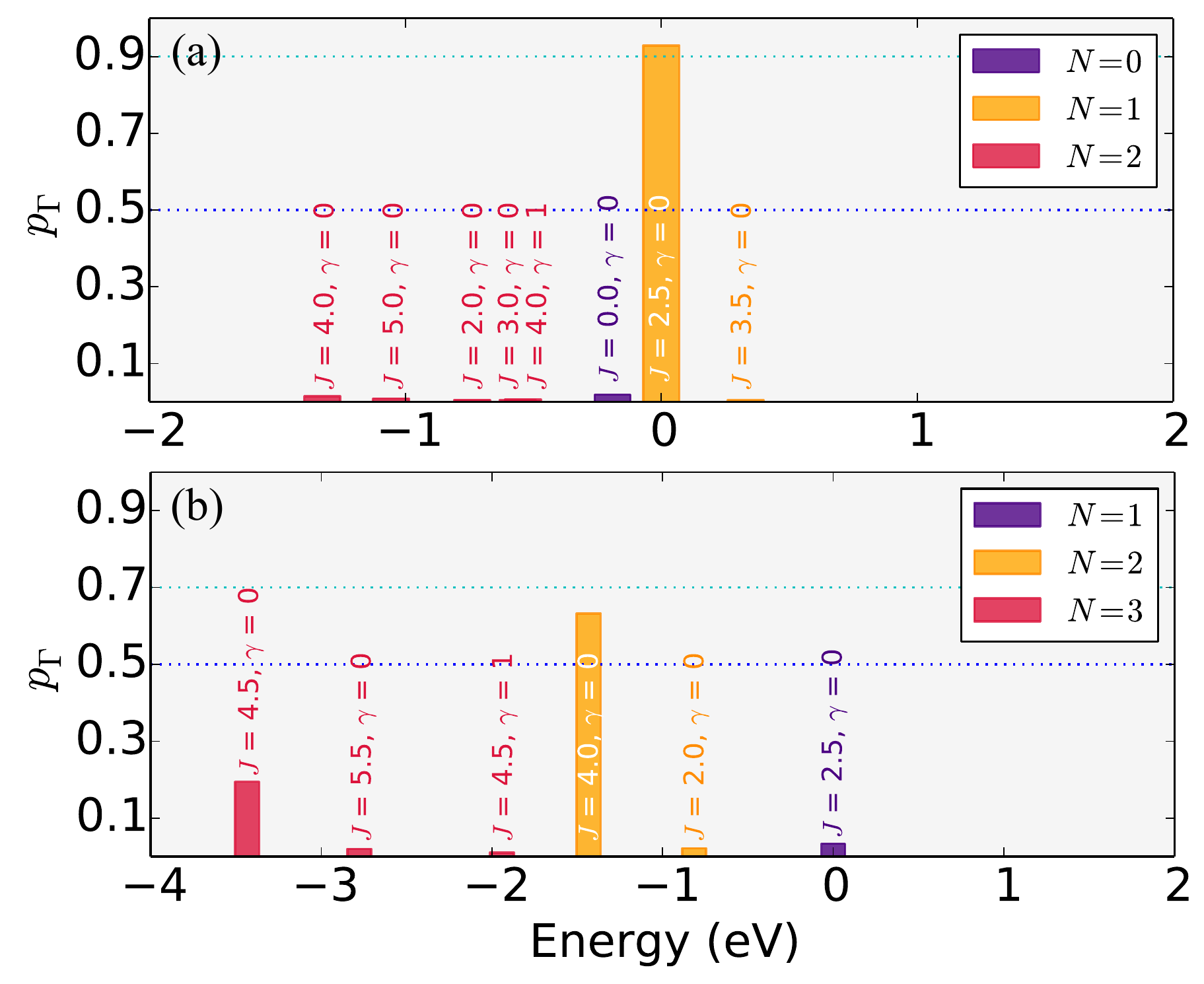}
\caption{(Color online). Valence state fluctuation in CeSb and USb by DFT + DMFT calculations. (a) Probabilities of the 4$f$ atomic eigenstates for CeSb. Here we used three good quantum numbers to label the atomic eigenstates. They are $N$ (total occupancy), $J$ (total angular momentum), and $\gamma$ ($\gamma$ stands for the rest of the atomic quantum numbers, such as $J_z$). (b) Probabilities of the 5$f$ atomic eigenstates for USb. \label{fig:prob}}
\end{figure}

\subsection{Photoemission spectroscopy}
To verify the reliability of our results, we first examine the calculated total density of states by referring to available experimental data. Figure~\ref{fig:tdos_exp} shows our calculated results and experimental density of states for CeSb[see Fig.~\ref{fig:tdos_exp}(a)] and USb[see Fig.~\ref{fig:tdos_exp}(b)]. It is apparent that almost no spectral weight exists near the Fermi level, revealing the localized 4$f$ states of CeSb. Furthermore, the calculated total density of states with a two-peak structure distribute around -4 eV $\sim$ 0 eV, which is consistent with ultraviolet photoemission spectra (UPS) data~\cite{PhysRevB.18.4433}. For USb, the most prominent feature is the significant quasi-particle resonance peak near the Fermi level, which indicates the itinerant behavior of 5$f$ electrons. Apart from that, a small shoulder peak about -2~eV is exactly reproduced, which agrees with UPS data~\cite{Reihl1982Electronic,Reihl1987Photoelectron}. The consistency between our computed density of states and available experiment data confirms the correctness and reasonability of our DFT + DMFT calculated results.

\subsection{Density of states and hybridization functions}
Here we discuss about the integrated spectral functions and hybridization functions of CeSb and USb. Figure~\ref{fig:tdos}(a1) and (a2) present total density of states $A(\omega)$ of CeSb and USb, respectively. It is evident that CeSb exhibits very weak quasi-particle resonance peak in the Fermi level, with one main peak centered at -2~eV, followed by the unoccupied 4$f^2$ multiplets around 2 eV $\sim$ 6 eV. Notice that the occupied states around the Fermi level are ascribed to 4$f_{5/2}$ states. The upper Hubbard band above the Fermi level is contributed by 4$f_{7/2}$ states. Combined with the partial density of states in Fig.~\ref{fig:tdos}(b1) and (b2), it is suggested that the low-lying 4$f_{5/2}$ states and high-lying 4$f_{7/2}$ states are split by the spin-orbital coupling with energy separation about 300~meV, which agrees with those observed in the other cerium-based heavy fermion compounds~\cite{PhysRevLett.108.016402,shim:1615}.
In Fig.~\ref{fig:tdos}(a2), a remarkable quasi-particle resonance peak emerges at the Fermi level, which comes from 5$f_{5/2}$ states of USb. Meanwhile, a pronounced peak centred at -2~eV roots from the hybridization between U-5$f$ bands and U-6$d$ bands. Moreover, the upper Hubbard band around 0 eV $\sim$ 6 eV from the unoccupied 5$f_{7/2}$ states forms a ``hump''. Then the hybridization strength between $f$ states and ligand $c$ bands are clearly characterized by the hybridization functions for $f_{5/2}$ and $f_{7/2}$ states[see Fig.~\ref{fig:tdos}(c1) and (c2)]. Obviously, $c-f$ hybridization of USb is much stronger than that of CeSb, demonstrating the localized 4$f$ states and partially itinerant 5$f$ ones.

\subsection{Momentum-resolved spectral functions}
In this section, we analyse the momentum-resolved spectral functions $A(\mathbf{k},\omega)$ of CeSb and USb[see Fig.~\ref{fig:akw}] along the high-symmetry line $X - \Gamma - X$ in the irreducible Brillouin zone[see Fig.~\ref{fig:tstruct}(b)]. As expected, the computed band structures reproduce the typical traits of experimental ARPES spectra~\cite{Takahashi199865,PhysRevB.61.15707,Takahashi2001High,Kumigashira1999High}. Figure~\ref{fig:akw}(a) plots the band structure of CeSb, which demonstrates missing 4$f$ bands and the prominent Sb-5$p$ states about -3 eV $\sim$ 0 eV. It is found that hole pockets at the $\Gamma$-point corresponding to Sb-5$p$ bands, and electron pockets at the $X$-point belonging to Ce-5$d$ states~\cite{PhysRevB.56.13654}.
On the other hand, the band structure of USb[see Fig.~\ref{fig:akw}(b)] displays two flat 5$f$ bands just above the Fermi level, hybridizing with conduction bands, which are in line with the itinerant-localized nature of 5$f$ states. It is discovered that an electron pocket just below the Fermi level at the $\Gamma$-point is attributed to U-6$d$ states and hole pockets around -3 eV $\sim$ -1 eV originate from Sb-5$p$ states. Beyond the previous theoretical studies~\cite{PhysRevB.22.645,JPSJ.64.2294,JPSJ.65.3394}, our calculation nicely captures the archetypical bands, further demonstrating the reliability of the DFT + DMFT method.

It is instructive to emphasize the representative band structures of CeSb and USb.
(i) The 4$f$ bands of CeSb are nearly invisible, in contrast, the 5$f$ bands of USb are clearly detected adjacent to the Fermi level. It is widely accepted that the position of $f$ bands relative to the Fermi level plays an important role in determining bonding and related physical properties.
(ii) The significant Sb-5$p$ states at the $\Gamma$-point hybridize with the partially occupied 4$f$ states, or totally occupied 5$f$ bands. Additionally, the band width of Sb-5$p$ states seems a bit smaller in CeSb than USb. (iii) The hybridization of U-5$f$ and U-6$d$ states right above the Fermi level manifests the itinerant 5$f$ states, embodying the itinerant-localized dual nature of USb. It is proposed that the correlation and itinerant degree of $f$ electrons are tightly related to the magnetism.

\subsection{Self-energy functions}

\begin{table}[t]
\caption{The effective electron mass $m^\star$ and quasi-particle weight $Z$ of $f_{5/2}$ and $f_{7/2}$ states for CeSb and USb. \label{tab:sig}}
\begin{ruledtabular}
\begin{tabular}{ccccc}
      & \multicolumn{2}{c}{$f_{5/2}$} & \multicolumn{2}{c}{$f_{7/2}$} \\
$R$Sb  & $m^{\star}/m_e$ & $Z$ & $m^{\star}/m_e$ & $Z$ \\
\hline
CeSb    & 43.425 & 0.023 & 1.001 & 0.999  \\
USb     & 02.476 & 0.404 & 3.086 & 0.324  \\
\end{tabular}
\end{ruledtabular}
\end{table}

Generally speaking, electronic correlations could be enclosed by self-energy functions. Figure~\ref{fig:tsigma} shows the Matsubara self-energy functions (only the imaginary parts at low frequency) for $f$ states of CeSb and USb. It is interesting to summarize the following features. Firstly, the self-energy function of 4$f_{5/2}$ state is convex at low frequency and 4$f_{7/2}$ component is concave. On the contrary, the low-frequency part of self-energy functions of 5$f_{5/2}$ state is obviously concave and 5$f_{7/2}$ state is slightly convex. Secondly, the low-energy scattering rate is described by the intercept in $y$-axis. It is found that the low-energy scattering rate of 4$f_{7/2}$ state approaches zero and is much smaller than the value of 4$f_{5/2}$ state. Thirdly, the low-energy scattering rate of 5$f$ states remains finite.

Then the quasi-particle weight $Z$ and effective electron mass $m^{\star}$ are evaluated via the following equation(Table~\ref{tab:sig}):
\begin{equation}
Z^{-1} = \frac{m^\star}{m_e} = 1 - \frac{\partial}{\partial \omega} \text{Re} \Sigma(\omega) \Big|_{\omega = 0}.
\end{equation}
Table~\ref{tab:sig} lists the computed effective electron mass $m^\star$ and quasi-particle weight $Z$ of $f_{5/2}$ and $f_{7/2}$ states for CeSb and USb. Evidently, the quasi-particle weight $Z$ of $4f_{7/2}$ states for CeSb is close to one, producing relatively small effective electron mass and suggesting the weakly correlated $4f_{7/2}$ states. Instead, the $4f_{5/2}$ bands are much more renormalized than the $4f_{7/2}$ bands with a large effective electron mass ($\approx 43.425 m_{e}$). Therefore the orbital differentiation of 4$f$ orbitals is notable in CeSb compared with the weakly renormalized 5$f$ bands in USb. Consequently, the 4$f$ electronic correlations are strongly orbital-selective and commonly exist in the other cerium-based heavy fermion compounds~\cite{PhysRevB.98.195102}.

\subsection{Valence state fluctuations}
Next let us concentrate on $f$ electronic configurations and valence state fluctuations for CeSb and USb. The calculated valence state histogram (or equivalently atomic eigenstate probability) $p_{\Gamma}$ for $f$ electrons is the direct output of the CT-HYB quantum impurity solver, which stands for the probability to find out a $f$ valence electron in a given atomic eigenstate $|\psi_\Gamma\rangle$ (labeled by perfect quantum numbers $N$, $J$ and the rest of atomic quantum numbers $\gamma$ as mentioned in Sec.~\ref{sec:method})~\cite{PhysRevB.75.155113}. If valence electrons only favor one or two dominant atomic eigenstates, it implies that the valence state fluctuation in such a system is weak or restricted. On the contrary, if valence electrons incline to wander in a large number of atomic eigenstates, the valence state fluctuation could be very strong.

Figure~\ref{fig:prob}(a) and (b) illustrate the calculated $f$ valence state histograms for CeSb and USb, respectively. For CeSb, the leading atomic eigenstate is $|N = 1, J = 2.5, \gamma = 0\rangle$ with the atomic eigenstate probability accounting for 93\%. At the same time, the atomic eigenstates probabilities for $|N = 0, J = 0.0, \gamma = 0\rangle$ and $|N = 1, J = 3.5, \gamma = 0\rangle$ are less than 1\%, which are nearly invisible[see Fig.~\ref{fig:prob}(a)]. It confirms the localized 4$f$ electrons in CeSb with valence state confined to the primary atomic eigenstate $|N = 1, J = 2.5, \gamma = 0\rangle$. Meanwhile, the corresponding valence state fluctuations are very weak. By summing up the atomic eigenstates probabilities $p_{\Gamma}$ with respect to $N$, we can derive the distribution of $f$ electronic configurations. It will provide further information about the $f$ valence state fluctuations and mixed-valence behaviors. In CeSb, the $4f^{1}$ configuration is predominant with its probability larger than 90\%, while those of the $4f^{0}$ and $4f^{2}$ configurations decline to less than $ 5\%$. Consequently, the mixed-valence behavior is suppressed. In comparison with CeSb, USb displays two competing atomic eigenstates $|N = 2, J = 4.0, \gamma = 0\rangle$ and $|N = 3, J = 4.5, \gamma = 0\rangle$ with the atomic eigenstate probabilities being 63\% and 19\%, respectively. In the meantime, the probabilities for the other atomic eigenstates are less than 3\%, which reveals that the 5$f$ electrons would fluctuate between the two principle atomic eigenstates and hybridize with conduction bands. Furthermore, the probabilities for the $5f^{2}$ and $5f^{3}$ configurations reach 69\% and 26\%, respectively, whereas those of $5f^{0}$ and $5f^{1}$ configurations are less than 4\%. Thus 5$f$ states are partially itinerant and valence state fluctuations become notable.

\section{Discussions\label{sec:dis}}

\emph{Similarities and differences between CeSb and USb.}
First of all, they crystalize in NaCl type of structure and their ground states are antiferromagnetic. 
Secondly, the electronic correlation of 4$f$ states is strongly orbital-dependent, while the 5$f$ states are moderately correlated. Thirdly, 4$f$ states tend to be localized with a small quasi-particle resonance peak near the Fermi level and show weak valence state fluctuations. In contrast, 5$f$ states display mixed-valence behavior with non-integer 5$f$ occupations. Lastly, the hybridization between 4$f$ states and partially occupied Sb-5$p$ states pushes the Sb-5$p$ band toward the Fermi level, changing the band structure and Fermi surface during the magnetic phase transition. Conversely, the completely occupied Sb-5$p$ states are further pushed down toward the lower binding energy after hybridizing with 5$f$ bands.

\emph{Unveiling the electronic correlation and magnetism.}
It is mentioned that the physical mechanism behind the ground states of CeSb and USb seem quite different. For CeSb, localized 4$f$ states are likely to form local moments, and the small crystal field splitting polarizes the spin along various directions to generate various magnetic phases~\cite{Busch1971Crystal}. It is anticipated that the hybridization between 4$f$ electrons and Sb-5$p$ states, as well as 4$f$ electronic correlations play a vital role in magnetic transition. 
However, it is another story for USb. The itinerant 5$f$ electrons locating above Sb-5$p$ states produce the extended spatial wave function and prefer to hybridize with totally occupied Sb-5$p$ states, which are pushed away from the Fermi level. Thus the $p-f$ mixing effect could not balance the energy cost during the magnetic transition~\cite{Takahashi2001High}. Therefore a credible model regarding the electronic correlation and crystal field splitting should be formulated to resolve such interesting problem. The present work is undertaken to interpret the long-standing issue.

\section{Conclusions\label{sec:summary}}
In summary, the subtle electronic structures of CeSb and USb have been comprehensively investigated using established DFT + DMFT method. The momentum-resolved spectral functions $A(\mathbf{k},\omega)$, total and $f$ partial density of states $A(\omega)$ and $A_{f}(\omega)$, hybridization functions, Matsubara self-energy functions, and $f$ valence state fluctuations are exhaustively studied. The calculated results not only accord with the available experimental data but also serve as critical predictions for future research. It is identified that the 4$f$ electrons in CeSb are generally localized while the 5$f$ states in USb exhibit itinerant-localized dual nature. 
Hence, tiny spectral weights near the Fermi level and weak valence state fluctuations are observed in CeSb. Instead a significant quasi-particle resonance peak appears around the Fermi level and mixed-valence behavior is remarkable in USb. Especially, it is suggested that 4$f$ states show orbital-dependent correlation, which is absent of 5$f$ states. Consequently, it is expected that the intrinsic electronic structure might lift the veil of the mysterious magnetism, which deserves further experimental and theoretical attention.

\acknowledgments
This work was supported by the Natural Science Foundation of China (No.~11704347 and No.~11904335), and the Science Challenge Project of China (No.~TZ2016004). The DFT + DMFT calculations were performed on the Sugon cluster (in the Institute of Physics, CAS, China).


\bibliography{RSb}

\begin{thebibliography}{73}%
\makeatletter
\providecommand \@ifxundefined [1]{%
 \@ifx{#1\undefined}
}%
\providecommand \@ifnum [1]{%
 \ifnum #1\expandafter \@firstoftwo
 \else \expandafter \@secondoftwo
 \fi
}%
\providecommand \@ifx [1]{%
 \ifx #1\expandafter \@firstoftwo
 \else \expandafter \@secondoftwo
 \fi
}%
\providecommand \natexlab [1]{#1}%
\providecommand \enquote  [1]{``#1''}%
\providecommand \bibnamefont  [1]{#1}%
\providecommand \bibfnamefont [1]{#1}%
\providecommand \citenamefont [1]{#1}%
\providecommand \href@noop [0]{\@secondoftwo}%
\providecommand \href [0]{\begingroup \@sanitize@url \@href}%
\providecommand \@href[1]{\@@startlink{#1}\@@href}%
\providecommand \@@href[1]{\endgroup#1\@@endlink}%
\providecommand \@sanitize@url [0]{\catcode `\\12\catcode `\$12\catcode
  `\&12\catcode `\#12\catcode `\^12\catcode `\_12\catcode `\%12\relax}%
\providecommand \@@startlink[1]{}%
\providecommand \@@endlink[0]{}%
\providecommand \url  [0]{\begingroup\@sanitize@url \@url }%
\providecommand \@url [1]{\endgroup\@href {#1}{\urlprefix }}%
\providecommand \urlprefix  [0]{URL }%
\providecommand \Eprint [0]{\href }%
\providecommand \doibase [0]{http://dx.doi.org/}%
\providecommand \selectlanguage [0]{\@gobble}%
\providecommand \bibinfo  [0]{\@secondoftwo}%
\providecommand \bibfield  [0]{\@secondoftwo}%
\providecommand \translation [1]{[#1]}%
\providecommand \BibitemOpen [0]{}%
\providecommand \bibitemStop [0]{}%
\providecommand \bibitemNoStop [0]{.\EOS\space}%
\providecommand \EOS [0]{\spacefactor3000\relax}%
\providecommand \BibitemShut  [1]{\csname bibitem#1\endcsname}%
\let\auto@bib@innerbib\@empty
\bibitem [{\citenamefont {Stewart}(1984)}]{RevModPhys.56.755}%
  \BibitemOpen
  \bibfield  {author} {\bibinfo {author} {\bibfnamefont {G.~R.}\ \bibnamefont
  {Stewart}},\ }\href {\doibase 10.1103/RevModPhys.56.755} {\bibfield
  {journal} {\bibinfo  {journal} {Rev. Mod. Phys.}\ }\textbf {\bibinfo {volume}
  {56}},\ \bibinfo {pages} {755} (\bibinfo {year} {1984})}\BibitemShut
  {NoStop}%
\bibitem [{\citenamefont {Sachdev}\ and\ \citenamefont
  {Keimer}(2011)}]{Sachdev2011Quantum}%
  \BibitemOpen
  \bibfield  {author} {\bibinfo {author} {\bibfnamefont {S.}~\bibnamefont
  {Sachdev}}\ and\ \bibinfo {author} {\bibfnamefont {B.}~\bibnamefont
  {Keimer}},\ }\href {\doibase 10.1063/1.3554314} {\bibfield  {journal}
  {\bibinfo  {journal} {Phys. Today}\ }\textbf {\bibinfo {volume} {64}},\
  \bibinfo {pages} {29} (\bibinfo {year} {2011})}\BibitemShut {NoStop}%
\bibitem [{\citenamefont {H\"alg}\ and\ \citenamefont
  {Furrer}(1986)}]{H1986Anisotropic}%
  \BibitemOpen
  \bibfield  {author} {\bibinfo {author} {\bibfnamefont {B.}~\bibnamefont
  {H\"alg}}\ and\ \bibinfo {author} {\bibfnamefont {A.}~\bibnamefont
  {Furrer}},\ }\href {\doibase 10.1103/PhysRevB.34.6258} {\bibfield  {journal}
  {\bibinfo  {journal} {Phys. Rev. B}\ }\textbf {\bibinfo {volume} {34}},\
  \bibinfo {pages} {6258} (\bibinfo {year} {1986})}\BibitemShut {NoStop}%
\bibitem [{\citenamefont {Lander}\ \emph {et~al.}(1976)\citenamefont {Lander},
  \citenamefont {Sparlin},\ and\ \citenamefont {Vogt}}]{Lander1976Neutron}%
  \BibitemOpen
  \bibfield  {author} {\bibinfo {author} {\bibfnamefont {G.~H.}\ \bibnamefont
  {Lander}}, \bibinfo {author} {\bibfnamefont {D.~M.}\ \bibnamefont {Sparlin}},
  \ and\ \bibinfo {author} {\bibfnamefont {O.}~\bibnamefont {Vogt}},\ }\href
  {\doibase 10.1103/PhysRevB.14.5035} {\bibfield  {journal} {\bibinfo
  {journal} {Phys. Rev. B}\ }\textbf {\bibinfo {volume} {14}},\ \bibinfo
  {pages} {5035} (\bibinfo {year} {1976})}\BibitemShut {NoStop}%
\bibitem [{\citenamefont {Lander}\ and\ \citenamefont
  {Stirling}(1980)}]{Lander1980Neutron}%
  \BibitemOpen
  \bibfield  {author} {\bibinfo {author} {\bibfnamefont {G.~H.}\ \bibnamefont
  {Lander}}\ and\ \bibinfo {author} {\bibfnamefont {W.~G.}\ \bibnamefont
  {Stirling}},\ }\href {\doibase 10.1103/PhysRevB.21.436} {\bibfield  {journal}
  {\bibinfo  {journal} {Phys. Rev. B}\ }\textbf {\bibinfo {volume} {21}},\
  \bibinfo {pages} {436} (\bibinfo {year} {1980})}\BibitemShut {NoStop}%
\bibitem [{\citenamefont {Lander}\ \emph {et~al.}(1979)\citenamefont {Lander},
  \citenamefont {Stirling},\ and\ \citenamefont {Vogt}}]{PhysRevLett.42.260}%
  \BibitemOpen
  \bibfield  {author} {\bibinfo {author} {\bibfnamefont {G.~H.}\ \bibnamefont
  {Lander}}, \bibinfo {author} {\bibfnamefont {W.~G.}\ \bibnamefont
  {Stirling}}, \ and\ \bibinfo {author} {\bibfnamefont {O.}~\bibnamefont
  {Vogt}},\ }\href {\doibase 10.1103/PhysRevLett.42.260} {\bibfield  {journal}
  {\bibinfo  {journal} {Phys. Rev. Lett.}\ }\textbf {\bibinfo {volume} {42}},\
  \bibinfo {pages} {260} (\bibinfo {year} {1979})}\BibitemShut {NoStop}%
\bibitem [{\citenamefont {Sarrao}\ \emph {et~al.}(2002)\citenamefont {Sarrao},
  \citenamefont {Morales}, \citenamefont {Thompson}, \citenamefont {Scott},
  \citenamefont {Stewart}, \citenamefont {Wastin}, \citenamefont {Rebizant},
  \citenamefont {Boulet}, \citenamefont {Colineau},\ and\ \citenamefont
  {Lander}}]{Sarrao2002}%
  \BibitemOpen
  \bibfield  {author} {\bibinfo {author} {\bibfnamefont {J.~L.}\ \bibnamefont
  {Sarrao}}, \bibinfo {author} {\bibfnamefont {L.~A.}\ \bibnamefont {Morales}},
  \bibinfo {author} {\bibfnamefont {J.~D.}\ \bibnamefont {Thompson}}, \bibinfo
  {author} {\bibfnamefont {B.~L.}\ \bibnamefont {Scott}}, \bibinfo {author}
  {\bibfnamefont {G.~R.}\ \bibnamefont {Stewart}}, \bibinfo {author}
  {\bibfnamefont {F.}~\bibnamefont {Wastin}}, \bibinfo {author} {\bibfnamefont
  {J.}~\bibnamefont {Rebizant}}, \bibinfo {author} {\bibfnamefont
  {P.}~\bibnamefont {Boulet}}, \bibinfo {author} {\bibfnamefont
  {E.}~\bibnamefont {Colineau}}, \ and\ \bibinfo {author} {\bibfnamefont
  {G.~H.}\ \bibnamefont {Lander}},\ }\href {\doibase 10.1038/nature01212}
  {\bibfield  {journal} {\bibinfo  {journal} {Nature}\ }\textbf {\bibinfo
  {volume} {420}},\ \bibinfo {pages} {297} (\bibinfo {year}
  {2002})}\BibitemShut {NoStop}%
\bibitem [{\citenamefont {Curro}\ \emph {et~al.}(2005)\citenamefont {Curro},
  \citenamefont {Caldwell}, \citenamefont {Bauer}, \citenamefont {Morales},
  \citenamefont {Graf}, \citenamefont {Bang}, \citenamefont {Balatsky},
  \citenamefont {Thompson},\ and\ \citenamefont {Sarrao}}]{Curro2005PuCoGa5}%
  \BibitemOpen
  \bibfield  {author} {\bibinfo {author} {\bibfnamefont {N.~J.}\ \bibnamefont
  {Curro}}, \bibinfo {author} {\bibfnamefont {T.}~\bibnamefont {Caldwell}},
  \bibinfo {author} {\bibfnamefont {E.~D.}\ \bibnamefont {Bauer}}, \bibinfo
  {author} {\bibfnamefont {L.~A.}\ \bibnamefont {Morales}}, \bibinfo {author}
  {\bibfnamefont {M.~J.}\ \bibnamefont {Graf}}, \bibinfo {author}
  {\bibfnamefont {Y.}~\bibnamefont {Bang}}, \bibinfo {author} {\bibfnamefont
  {A.~V.}\ \bibnamefont {Balatsky}}, \bibinfo {author} {\bibfnamefont {J.~D.}\
  \bibnamefont {Thompson}}, \ and\ \bibinfo {author} {\bibfnamefont {J.~L.}\
  \bibnamefont {Sarrao}},\ }\href {\doibase 10.1038/nature03428} {\bibfield
  {journal} {\bibinfo  {journal} {Nature}\ }\textbf {\bibinfo {volume} {434}},\
  \bibinfo {pages} {622} (\bibinfo {year} {2005})}\BibitemShut {NoStop}%
\bibitem [{\citenamefont {Moore}\ and\ \citenamefont {van~der
  Laan}(2009)}]{RevModPhys.81.235}%
  \BibitemOpen
  \bibfield  {author} {\bibinfo {author} {\bibfnamefont {K.~T.}\ \bibnamefont
  {Moore}}\ and\ \bibinfo {author} {\bibfnamefont {G.}~\bibnamefont {van~der
  Laan}},\ }\href {\doibase 10.1103/RevModPhys.81.235} {\bibfield  {journal}
  {\bibinfo  {journal} {Rev. Mod. Phys.}\ }\textbf {\bibinfo {volume} {81}},\
  \bibinfo {pages} {235} (\bibinfo {year} {2009})}\BibitemShut {NoStop}%
\bibitem [{\citenamefont {Bartholin}\ \emph {et~al.}(1977)\citenamefont
  {Bartholin}, \citenamefont {Florence}, \citenamefont {Parisot}, \citenamefont
  {Paureau},\ and\ \citenamefont {Vogt}}]{Bartholin1977}%
  \BibitemOpen
  \bibfield  {author} {\bibinfo {author} {\bibfnamefont {H.}~\bibnamefont
  {Bartholin}}, \bibinfo {author} {\bibfnamefont {D.}~\bibnamefont {Florence}},
  \bibinfo {author} {\bibfnamefont {G.}~\bibnamefont {Parisot}}, \bibinfo
  {author} {\bibfnamefont {J.}~\bibnamefont {Paureau}}, \ and\ \bibinfo
  {author} {\bibfnamefont {O.}~\bibnamefont {Vogt}},\ }\href {\doibase
  10.1016/0375-9601(77)90316-4} {\bibfield  {journal} {\bibinfo  {journal}
  {Phys. Lett. A}\ }\textbf {\bibinfo {volume} {60}},\ \bibinfo {pages} {47}
  (\bibinfo {year} {1977})}\BibitemShut {NoStop}%
\bibitem [{\citenamefont {Kuznietz}\ \emph {et~al.}(1969)\citenamefont
  {Kuznietz}, \citenamefont {Lander},\ and\ \citenamefont
  {Campos}}]{Kuznietz1969}%
  \BibitemOpen
  \bibfield  {author} {\bibinfo {author} {\bibfnamefont {M.}~\bibnamefont
  {Kuznietz}}, \bibinfo {author} {\bibfnamefont {G.~H.}\ \bibnamefont
  {Lander}}, \ and\ \bibinfo {author} {\bibfnamefont {F.~P.}\ \bibnamefont
  {Campos}},\ }\href {\doibase 10.1016/0022-3697(69)90229-7} {\bibfield
  {journal} {\bibinfo  {journal} {J. Phys. Chem. Solids}\ }\textbf {\bibinfo
  {volume} {30}},\ \bibinfo {pages} {1642} (\bibinfo {year}
  {1969})}\BibitemShut {NoStop}%
\bibitem [{\citenamefont {Rossat-Migno}\ \emph {et~al.}(1980)\citenamefont
  {Rossat-Migno}, \citenamefont {Burlet}, \citenamefont {Quezela},\ and\
  \citenamefont {Vogtb}}]{Rossat1980Magnetic}%
  \BibitemOpen
  \bibfield  {author} {\bibinfo {author} {\bibfnamefont {J.}~\bibnamefont
  {Rossat-Migno}}, \bibinfo {author} {\bibfnamefont {P.}~\bibnamefont
  {Burlet}}, \bibinfo {author} {\bibfnamefont {S.}~\bibnamefont {Quezela}}, \
  and\ \bibinfo {author} {\bibfnamefont {O.}~\bibnamefont {Vogtb}},\ }\href
  {\doibase 10.1016/0378-4363(80)90165-5} {\bibfield  {journal} {\bibinfo
  {journal} {Physica B+C}\ }\textbf {\bibinfo {volume} {102}},\ \bibinfo
  {pages} {237} (\bibinfo {year} {1980})}\BibitemShut {NoStop}%
\bibitem [{\citenamefont {Hulliger}\ \emph {et~al.}(1975)\citenamefont
  {Hulliger}, \citenamefont {Landolt}, \citenamefont {Ott},\ and\ \citenamefont
  {Schmelczer}}]{Hulliger1975Low}%
  \BibitemOpen
  \bibfield  {author} {\bibinfo {author} {\bibfnamefont {F.}~\bibnamefont
  {Hulliger}}, \bibinfo {author} {\bibfnamefont {M.}~\bibnamefont {Landolt}},
  \bibinfo {author} {\bibfnamefont {H.~R.}\ \bibnamefont {Ott}}, \ and\
  \bibinfo {author} {\bibfnamefont {R.}~\bibnamefont {Schmelczer}},\ }\href
  {\doibase 10.1007/BF00117797} {\bibfield  {journal} {\bibinfo  {journal} {J.
  Low Temp. Phys.}\ }\textbf {\bibinfo {volume} {20}},\ \bibinfo {pages} {269}
  (\bibinfo {year} {1975})}\BibitemShut {NoStop}%
\bibitem [{\citenamefont {Fischer}\ \emph {et~al.}(1978)\citenamefont
  {Fischer}, \citenamefont {Meier}, \citenamefont {Lebech}, \citenamefont
  {Rainford},\ and\ \citenamefont {Vogt}}]{Fischer1978Magnetic}%
  \BibitemOpen
  \bibfield  {author} {\bibinfo {author} {\bibfnamefont {P.}~\bibnamefont
  {Fischer}}, \bibinfo {author} {\bibfnamefont {G.}~\bibnamefont {Meier}},
  \bibinfo {author} {\bibfnamefont {B.}~\bibnamefont {Lebech}}, \bibinfo
  {author} {\bibfnamefont {B.~D.}\ \bibnamefont {Rainford}}, \ and\ \bibinfo
  {author} {\bibfnamefont {O.}~\bibnamefont {Vogt}},\ }\href {\doibase
  10.1088/0022-3719/11/2/018} {\bibfield  {journal} {\bibinfo  {journal} {J.
  Phys. C: Solid State Phys.}\ }\textbf {\bibinfo {volume} {11}},\ \bibinfo
  {pages} {345} (\bibinfo {year} {1978})}\BibitemShut {NoStop}%
\bibitem [{\citenamefont {Meier}\ \emph {et~al.}(1978)\citenamefont {Meier},
  \citenamefont {Fischer}, \citenamefont {Halg}, \citenamefont {Lebech},
  \citenamefont {Rainford},\ and\ \citenamefont {Vogt}}]{Meier1978Magnetic}%
  \BibitemOpen
  \bibfield  {author} {\bibinfo {author} {\bibfnamefont {G.}~\bibnamefont
  {Meier}}, \bibinfo {author} {\bibfnamefont {P.}~\bibnamefont {Fischer}},
  \bibinfo {author} {\bibfnamefont {W.}~\bibnamefont {Halg}}, \bibinfo {author}
  {\bibfnamefont {B.}~\bibnamefont {Lebech}}, \bibinfo {author} {\bibfnamefont
  {B.~D.}\ \bibnamefont {Rainford}}, \ and\ \bibinfo {author} {\bibfnamefont
  {O.}~\bibnamefont {Vogt}},\ }\href {\doibase 10.1088/0022-3719/11/6/023}
  {\bibfield  {journal} {\bibinfo  {journal} {J. Phys. C: Solid State Phys.}\
  }\textbf {\bibinfo {volume} {11}},\ \bibinfo {pages} {1173} (\bibinfo {year}
  {1978})}\BibitemShut {NoStop}%
\bibitem [{\citenamefont {Furrer}\ \emph {et~al.}(1979)\citenamefont {Furrer},
  \citenamefont {H\"alg}, \citenamefont {Heer},\ and\ \citenamefont
  {Vogt}}]{Furrer1979Magnetic}%
  \BibitemOpen
  \bibfield  {author} {\bibinfo {author} {\bibfnamefont {A.}~\bibnamefont
  {Furrer}}, \bibinfo {author} {\bibfnamefont {W.}~\bibnamefont {H\"alg}},
  \bibinfo {author} {\bibfnamefont {H.}~\bibnamefont {Heer}}, \ and\ \bibinfo
  {author} {\bibfnamefont {O.}~\bibnamefont {Vogt}},\ }\href {\doibase
  10.1063/1.327101} {\bibfield  {journal} {\bibinfo  {journal} {J. Appl.
  Phys.}\ }\textbf {\bibinfo {volume} {50}},\ \bibinfo {pages} {2040} (\bibinfo
  {year} {1979})}\BibitemShut {NoStop}%
\bibitem [{\citenamefont {Lebech}\ and\ \citenamefont
  {Bente}(1981)}]{Lebech1981Neutron}%
  \BibitemOpen
  \bibfield  {author} {\bibinfo {author} {\bibnamefont {Lebech}}\ and\ \bibinfo
  {author} {\bibnamefont {Bente}},\ }\href {\doibase 10.1063/1.329599}
  {\bibfield  {journal} {\bibinfo  {journal} {J. Appl. Phys.}\ }\textbf
  {\bibinfo {volume} {52}},\ \bibinfo {pages} {2019} (\bibinfo {year}
  {1981})}\BibitemShut {NoStop}%
\bibitem [{\citenamefont {Rossat-Mignod}\ \emph {et~al.}(1985)\citenamefont
  {Rossat-Mignod}, \citenamefont {Effantin}, \citenamefont {Burlet},
  \citenamefont {Chattopadhyay}, \citenamefont {Regnault}, \citenamefont
  {Bartholin}, \citenamefont {Vettier}, \citenamefont {Vogt}, \citenamefont
  {Ravot},\ and\ \citenamefont {Achart}}]{Rossat1985Neutron}%
  \BibitemOpen
  \bibfield  {author} {\bibinfo {author} {\bibfnamefont {J.}~\bibnamefont
  {Rossat-Mignod}}, \bibinfo {author} {\bibfnamefont {J.}~\bibnamefont
  {Effantin}}, \bibinfo {author} {\bibfnamefont {P.}~\bibnamefont {Burlet}},
  \bibinfo {author} {\bibfnamefont {T.}~\bibnamefont {Chattopadhyay}}, \bibinfo
  {author} {\bibfnamefont {L.}~\bibnamefont {Regnault}}, \bibinfo {author}
  {\bibfnamefont {H.}~\bibnamefont {Bartholin}}, \bibinfo {author}
  {\bibfnamefont {C.}~\bibnamefont {Vettier}}, \bibinfo {author} {\bibfnamefont
  {O.}~\bibnamefont {Vogt}}, \bibinfo {author} {\bibfnamefont {D.}~\bibnamefont
  {Ravot}}, \ and\ \bibinfo {author} {\bibfnamefont {J.}~\bibnamefont
  {Achart}},\ }\href {\doibase 0.1016/0304-8853(85)90235-5} {\bibfield
  {journal} {\bibinfo  {journal} {J. Magn. Magn. Mater.}\ }\textbf {\bibinfo
  {volume} {52}},\ \bibinfo {pages} {111} (\bibinfo {year} {1985})}\BibitemShut
  {NoStop}%
\bibitem [{\citenamefont {Chattopadhyay}\ \emph {et~al.}(1987)\citenamefont
  {Chattopadhyay}, \citenamefont {Burlet}, \citenamefont {Rossat-Mignod},
  \citenamefont {Bartholin}, \citenamefont {Vettier},\ and\ \citenamefont
  {Vogt}}]{Chattopadhyay1994High}%
  \BibitemOpen
  \bibfield  {author} {\bibinfo {author} {\bibfnamefont {T.}~\bibnamefont
  {Chattopadhyay}}, \bibinfo {author} {\bibfnamefont {P.}~\bibnamefont
  {Burlet}}, \bibinfo {author} {\bibfnamefont {J.}~\bibnamefont
  {Rossat-Mignod}}, \bibinfo {author} {\bibfnamefont {H.}~\bibnamefont
  {Bartholin}}, \bibinfo {author} {\bibfnamefont {C.}~\bibnamefont {Vettier}},
  \ and\ \bibinfo {author} {\bibfnamefont {O.}~\bibnamefont {Vogt}},\ }\href
  {\doibase 10.1016/0304-8853(87)90519-1} {\bibfield  {journal} {\bibinfo
  {journal} {J. Magn. Magn. Mater.}\ }\textbf {\bibinfo {volume} {63}},\
  \bibinfo {pages} {52} (\bibinfo {year} {1987})}\BibitemShut {NoStop}%
\bibitem [{\citenamefont {Vogt}\ and\ \citenamefont
  {Mattenberger}(1995)}]{Vogt1995The}%
  \BibitemOpen
  \bibfield  {author} {\bibinfo {author} {\bibfnamefont {O.}~\bibnamefont
  {Vogt}}\ and\ \bibinfo {author} {\bibfnamefont {K.}~\bibnamefont
  {Mattenberger}},\ }\href {\doibase 10.1016/0921-4526(95)00022-2} {\bibfield
  {journal} {\bibinfo  {journal} {Phys. B Condens. Matter}\ }\textbf {\bibinfo
  {volume} {215}},\ \bibinfo {pages} {22} (\bibinfo {year} {1995})}\BibitemShut
  {NoStop}%
\bibitem [{\citenamefont {Kohgi}\ \emph {et~al.}(2000)\citenamefont {Kohgi},
  \citenamefont {Iwasa},\ and\ \citenamefont {Osakabe}}]{Kohgi2000Physics}%
  \BibitemOpen
  \bibfield  {author} {\bibinfo {author} {\bibfnamefont {M.}~\bibnamefont
  {Kohgi}}, \bibinfo {author} {\bibfnamefont {K.}~\bibnamefont {Iwasa}}, \ and\
  \bibinfo {author} {\bibfnamefont {T.}~\bibnamefont {Osakabe}},\ }\href
  {\doibase 10.1016/S0921-4526(99)01051-0} {\bibfield  {journal} {\bibinfo
  {journal} {Phys. B Condens. Matter}\ }\textbf {\bibinfo {volume} {281}},\
  \bibinfo {pages} {417} (\bibinfo {year} {2000})}\BibitemShut {NoStop}%
\bibitem [{\citenamefont {Busch}\ \emph {et~al.}(1971)\citenamefont {Busch},
  \citenamefont {Stutius},\ and\ \citenamefont {Vogt}}]{Busch1971Crystal}%
  \BibitemOpen
  \bibfield  {author} {\bibinfo {author} {\bibfnamefont {G.}~\bibnamefont
  {Busch}}, \bibinfo {author} {\bibfnamefont {W.}~\bibnamefont {Stutius}}, \
  and\ \bibinfo {author} {\bibfnamefont {O.}~\bibnamefont {Vogt}},\ }\href
  {\doibase 10.1063/1.1660314} {\bibfield  {journal} {\bibinfo  {journal} {J.
  Appl. Phys.}\ }\textbf {\bibinfo {volume} {42}},\ \bibinfo {pages} {1493}
  (\bibinfo {year} {1971})}\BibitemShut {NoStop}%
\bibitem [{\citenamefont {Kumigashira}\ \emph {et~al.}(1997)\citenamefont
  {Kumigashira}, \citenamefont {Kim}, \citenamefont {Ashihara}, \citenamefont
  {Chainani}, \citenamefont {Yokoya}, \citenamefont {Takahashi}, \citenamefont
  {Uesawa},\ and\ \citenamefont {Suzuki}}]{PhysRevB.56.13654}%
  \BibitemOpen
  \bibfield  {author} {\bibinfo {author} {\bibfnamefont {H.}~\bibnamefont
  {Kumigashira}}, \bibinfo {author} {\bibfnamefont {H.-D.}\ \bibnamefont
  {Kim}}, \bibinfo {author} {\bibfnamefont {A.}~\bibnamefont {Ashihara}},
  \bibinfo {author} {\bibfnamefont {A.}~\bibnamefont {Chainani}}, \bibinfo
  {author} {\bibfnamefont {T.}~\bibnamefont {Yokoya}}, \bibinfo {author}
  {\bibfnamefont {T.}~\bibnamefont {Takahashi}}, \bibinfo {author}
  {\bibfnamefont {A.}~\bibnamefont {Uesawa}}, \ and\ \bibinfo {author}
  {\bibfnamefont {T.}~\bibnamefont {Suzuki}},\ }\href {\doibase
  10.1103/PhysRevB.56.13654} {\bibfield  {journal} {\bibinfo  {journal} {Phys.
  Rev. B}\ }\textbf {\bibinfo {volume} {56}},\ \bibinfo {pages} {13654}
  (\bibinfo {year} {1997})}\BibitemShut {NoStop}%
\bibitem [{\citenamefont {Takahashi}\ \emph {et~al.}(1998)\citenamefont
  {Takahashi}, \citenamefont {Kumigashira}, \citenamefont {Ito}, \citenamefont
  {Ashihara}, \citenamefont {Kim}, \citenamefont {Aoki}, \citenamefont
  {Ochiai},\ and\ \citenamefont {Suzuki}}]{Takahashi199865}%
  \BibitemOpen
  \bibfield  {author} {\bibinfo {author} {\bibfnamefont {T.}~\bibnamefont
  {Takahashi}}, \bibinfo {author} {\bibfnamefont {H.}~\bibnamefont
  {Kumigashira}}, \bibinfo {author} {\bibfnamefont {T.}~\bibnamefont {Ito}},
  \bibinfo {author} {\bibfnamefont {A.}~\bibnamefont {Ashihara}}, \bibinfo
  {author} {\bibfnamefont {H.}~\bibnamefont {Kim}}, \bibinfo {author}
  {\bibfnamefont {H.}~\bibnamefont {Aoki}}, \bibinfo {author} {\bibfnamefont
  {A.}~\bibnamefont {Ochiai}}, \ and\ \bibinfo {author} {\bibfnamefont
  {T.}~\bibnamefont {Suzuki}},\ }\href {\doibase
  http://dx.doi.org/10.1016/S0368-2048(98)00102-9} {\bibfield  {journal}
  {\bibinfo  {journal} {J. Electron Spectrosc. Relat. Phenom.}\ }\textbf
  {\bibinfo {volume} {92}},\ \bibinfo {pages} {65} (\bibinfo {year}
  {1998})}\BibitemShut {NoStop}%
\bibitem [{\citenamefont {Ito}\ \emph {et~al.}(2004)\citenamefont {Ito},
  \citenamefont {Kimura},\ and\ \citenamefont {Kitazawa}}]{Ito2004Para}%
  \BibitemOpen
  \bibfield  {author} {\bibinfo {author} {\bibfnamefont {T.}~\bibnamefont
  {Ito}}, \bibinfo {author} {\bibfnamefont {S.}~\bibnamefont {Kimura}}, \ and\
  \bibinfo {author} {\bibfnamefont {H.}~\bibnamefont {Kitazawa}},\ }\href
  {\doibase 10.1016/j.physb.2004.06.022} {\bibfield  {journal} {\bibinfo
  {journal} {Phys. B Condens. Matter}\ }\textbf {\bibinfo {volume} {351}},\
  \bibinfo {pages} {268} (\bibinfo {year} {2004})}\BibitemShut {NoStop}%
\bibitem [{\citenamefont {Takayama}\ \emph {et~al.}(2011)\citenamefont
  {Takayama}, \citenamefont {Souma}, \citenamefont {Sato}, \citenamefont
  {Arakane},\ and\ \citenamefont {Takahashi}}]{Takayama2011Magnetic}%
  \BibitemOpen
  \bibfield  {author} {\bibinfo {author} {\bibfnamefont {A.}~\bibnamefont
  {Takayama}}, \bibinfo {author} {\bibfnamefont {S.}~\bibnamefont {Souma}},
  \bibinfo {author} {\bibfnamefont {T.}~\bibnamefont {Sato}}, \bibinfo {author}
  {\bibfnamefont {T.}~\bibnamefont {Arakane}}, \ and\ \bibinfo {author}
  {\bibfnamefont {T.}~\bibnamefont {Takahashi}},\ }\href {\doibase
  10.1143/JPSJ.78.073702} {\bibfield  {journal} {\bibinfo  {journal} {J. Phys.
  Soc. Jpn.}\ }\textbf {\bibinfo {volume} {78}},\ \bibinfo {pages} {3702}
  (\bibinfo {year} {2011})}\BibitemShut {NoStop}%
\bibitem [{\citenamefont {Takahashi}(2001)}]{Takahashi2001High}%
  \BibitemOpen
  \bibfield  {author} {\bibinfo {author} {\bibfnamefont {T.}~\bibnamefont
  {Takahashi}},\ }\href {\doibase 10.1016/s0368-2048(01)00255-9} {\bibfield
  {journal} {\bibinfo  {journal} {J. Electron Spectrosc. Relat. Phenom.}\
  }\textbf {\bibinfo {volume} {117}},\ \bibinfo {pages} {311} (\bibinfo {year}
  {2001})}\BibitemShut {NoStop}%
\bibitem [{\citenamefont {Duan}\ \emph {et~al.}(2007)\citenamefont {Duan},
  \citenamefont {Sabirianov}, \citenamefont {Mei}, \citenamefont {Dowben},
  \citenamefont {Jaswal},\ and\ \citenamefont {Tsymbal}}]{Duan2007Electronic}%
  \BibitemOpen
  \bibfield  {author} {\bibinfo {author} {\bibfnamefont {C.-G.}\ \bibnamefont
  {Duan}}, \bibinfo {author} {\bibfnamefont {R.~F.}\ \bibnamefont
  {Sabirianov}}, \bibinfo {author} {\bibfnamefont {W.~N.}\ \bibnamefont {Mei}},
  \bibinfo {author} {\bibfnamefont {P.~A.}\ \bibnamefont {Dowben}}, \bibinfo
  {author} {\bibfnamefont {S.~S.}\ \bibnamefont {Jaswal}}, \ and\ \bibinfo
  {author} {\bibfnamefont {E.~Y.}\ \bibnamefont {Tsymbal}},\ }\href {\doibase
  10.1088/0953-8984/19/31/315220} {\bibfield  {journal} {\bibinfo  {journal}
  {J. Phys.: Condens. Matter}\ }\textbf {\bibinfo {volume} {19}},\ \bibinfo
  {pages} {315220} (\bibinfo {year} {2007})}\BibitemShut {NoStop}%
\bibitem [{\citenamefont {Sakai}\ and\ \citenamefont
  {Shimizu}(2007{\natexlab{a}})}]{JPSJ.76.044707}%
  \BibitemOpen
  \bibfield  {author} {\bibinfo {author} {\bibfnamefont {O.}~\bibnamefont
  {Sakai}}\ and\ \bibinfo {author} {\bibfnamefont {Y.}~\bibnamefont
  {Shimizu}},\ }\href {\doibase 10.1143/JPSJ.76.044707} {\bibfield  {journal}
  {\bibinfo  {journal} {J. Phys. Soc. Jpn.}\ }\textbf {\bibinfo {volume}
  {76}},\ \bibinfo {pages} {044707} (\bibinfo {year}
  {2007}{\natexlab{a}})}\BibitemShut {NoStop}%
\bibitem [{\citenamefont {Sakai}\ and\ \citenamefont
  {Shimizu}(2007{\natexlab{b}})}]{Sakai2007}%
  \BibitemOpen
  \bibfield  {author} {\bibinfo {author} {\bibfnamefont {O.}~\bibnamefont
  {Sakai}}\ and\ \bibinfo {author} {\bibfnamefont {Y.}~\bibnamefont
  {Shimizu}},\ }\href {\doibase 10.1088/0953-8984/19/36/365213} {\bibfield
  {journal} {\bibinfo  {journal} {J. Phys.: Condens. Matter}\ }\textbf
  {\bibinfo {volume} {19}},\ \bibinfo {pages} {365213} (\bibinfo {year}
  {2007}{\natexlab{b}})}\BibitemShut {NoStop}%
\bibitem [{\citenamefont {Litsarev}\ \emph {et~al.}(2012)\citenamefont
  {Litsarev}, \citenamefont {Di~Marco}, \citenamefont {Thunstr\"om},\ and\
  \citenamefont {Eriksson}}]{PhysRevB.86.115116}%
  \BibitemOpen
  \bibfield  {author} {\bibinfo {author} {\bibfnamefont {M.~S.}\ \bibnamefont
  {Litsarev}}, \bibinfo {author} {\bibfnamefont {I.}~\bibnamefont {Di~Marco}},
  \bibinfo {author} {\bibfnamefont {P.}~\bibnamefont {Thunstr\"om}}, \ and\
  \bibinfo {author} {\bibfnamefont {O.}~\bibnamefont {Eriksson}},\ }\href
  {\doibase 10.1103/PhysRevB.86.115116} {\bibfield  {journal} {\bibinfo
  {journal} {Phys. Rev. B}\ }\textbf {\bibinfo {volume} {86}},\ \bibinfo
  {pages} {115116} (\bibinfo {year} {2012})}\BibitemShut {NoStop}%
\bibitem [{\citenamefont {Sakai}\ \emph {et~al.}(2012)\citenamefont {Sakai},
  \citenamefont {Shimizu},\ and\ \citenamefont {Kaneta}}]{Sakai2012}%
  \BibitemOpen
  \bibfield  {author} {\bibinfo {author} {\bibfnamefont {O.}~\bibnamefont
  {Sakai}}, \bibinfo {author} {\bibfnamefont {Y.}~\bibnamefont {Shimizu}}, \
  and\ \bibinfo {author} {\bibfnamefont {Y.}~\bibnamefont {Kaneta}},\ }\href
  {\doibase 10.1143/JPSJ.74.2517} {\bibfield  {journal} {\bibinfo  {journal}
  {J. Phys. Soc. Jpn.}\ }\textbf {\bibinfo {volume} {74}},\ \bibinfo {pages}
  {2517} (\bibinfo {year} {2012})}\BibitemShut {NoStop}%
\bibitem [{\citenamefont {Rudigier}\ \emph {et~al.}(1983)\citenamefont
  {Rudigier}, \citenamefont {Fierz}, \citenamefont {Ott},\ and\ \citenamefont
  {Vogt}}]{Rudigier1983Low}%
  \BibitemOpen
  \bibfield  {author} {\bibinfo {author} {\bibfnamefont {H.}~\bibnamefont
  {Rudigier}}, \bibinfo {author} {\bibfnamefont {C.}~\bibnamefont {Fierz}},
  \bibinfo {author} {\bibfnamefont {H.~R.}\ \bibnamefont {Ott}}, \ and\
  \bibinfo {author} {\bibfnamefont {O.}~\bibnamefont {Vogt}},\ }\href {\doibase
  10.1016/0038-1098(83)90070-4} {\bibfield  {journal} {\bibinfo  {journal}
  {Solid State Commun.}\ }\textbf {\bibinfo {volume} {47}},\ \bibinfo {pages}
  {803} (\bibinfo {year} {1983})}\BibitemShut {NoStop}%
\bibitem [{\citenamefont {Rudigier}\ \emph {et~al.}(1985)\citenamefont
  {Rudigier}, \citenamefont {Ott},\ and\ \citenamefont
  {Vogt}}]{PhysRevB.32.4584}%
  \BibitemOpen
  \bibfield  {author} {\bibinfo {author} {\bibfnamefont {H.}~\bibnamefont
  {Rudigier}}, \bibinfo {author} {\bibfnamefont {H.~R.}\ \bibnamefont {Ott}}, \
  and\ \bibinfo {author} {\bibfnamefont {O.}~\bibnamefont {Vogt}},\ }\href
  {\doibase 10.1103/PhysRevB.32.4584} {\bibfield  {journal} {\bibinfo
  {journal} {Phys. Rev. B}\ }\textbf {\bibinfo {volume} {32}},\ \bibinfo
  {pages} {4584} (\bibinfo {year} {1985})}\BibitemShut {NoStop}%
\bibitem [{\citenamefont {Schoenes}\ \emph {et~al.}(1984)\citenamefont
  {Schoenes}, \citenamefont {Frick},\ and\ \citenamefont
  {Vogt}}]{PhysRevB.30.6578}%
  \BibitemOpen
  \bibfield  {author} {\bibinfo {author} {\bibfnamefont {J.}~\bibnamefont
  {Schoenes}}, \bibinfo {author} {\bibfnamefont {B.}~\bibnamefont {Frick}}, \
  and\ \bibinfo {author} {\bibfnamefont {O.}~\bibnamefont {Vogt}},\ }\href
  {\doibase 10.1103/PhysRevB.30.6578} {\bibfield  {journal} {\bibinfo
  {journal} {Phys. Rev. B}\ }\textbf {\bibinfo {volume} {30}},\ \bibinfo
  {pages} {6578} (\bibinfo {year} {1984})}\BibitemShut {NoStop}%
\bibitem [{\citenamefont {Lander}\ and\ \citenamefont
  {Burlet}(1995)}]{Lander1995On}%
  \BibitemOpen
  \bibfield  {author} {\bibinfo {author} {\bibfnamefont {G.~H.}\ \bibnamefont
  {Lander}}\ and\ \bibinfo {author} {\bibfnamefont {P.}~\bibnamefont
  {Burlet}},\ }\href {\doibase 10.1016/0921-4526(95)00021-z} {\bibfield
  {journal} {\bibinfo  {journal} {Phys. B Condens. Matter}\ }\textbf {\bibinfo
  {volume} {215}},\ \bibinfo {pages} {7} (\bibinfo {year} {1995})}\BibitemShut
  {NoStop}%
\bibitem [{\citenamefont {Lander}\ and\ \citenamefont
  {Shapiro}(1995)}]{Lander1995Possible}%
  \BibitemOpen
  \bibfield  {author} {\bibinfo {author} {\bibfnamefont {G.~H.}\ \bibnamefont
  {Lander}}\ and\ \bibinfo {author} {\bibfnamefont {S.~M.}\ \bibnamefont
  {Shapiro}},\ }\href {\doibase 10.1016/0921-4526(95)00080-s} {\bibfield
  {journal} {\bibinfo  {journal} {Phys. B Condens. Matter}\ }\textbf {\bibinfo
  {volume} {213}},\ \bibinfo {pages} {125} (\bibinfo {year}
  {1995})}\BibitemShut {NoStop}%
\bibitem [{\citenamefont {Grunzweig-Genossar}\ \emph
  {et~al.}(1968)\citenamefont {Grunzweig-Genossar}, \citenamefont {Kuznietz},\
  and\ \citenamefont {Friedman}}]{PhysRev.173.562}%
  \BibitemOpen
  \bibfield  {author} {\bibinfo {author} {\bibfnamefont {J.}~\bibnamefont
  {Grunzweig-Genossar}}, \bibinfo {author} {\bibfnamefont {M.}~\bibnamefont
  {Kuznietz}}, \ and\ \bibinfo {author} {\bibfnamefont {F.}~\bibnamefont
  {Friedman}},\ }\href {\doibase 10.1103/PhysRev.173.562} {\bibfield  {journal}
  {\bibinfo  {journal} {Phys. Rev.}\ }\textbf {\bibinfo {volume} {173}},\
  \bibinfo {pages} {562} (\bibinfo {year} {1968})}\BibitemShut {NoStop}%
\bibitem [{\citenamefont {Hälg}\ and\ \citenamefont
  {Vogt}(1985)}]{H1985Magnetic}%
  \BibitemOpen
  \bibfield  {author} {\bibinfo {author} {\bibfnamefont {B.}~\bibnamefont
  {Hälg}}\ and\ \bibinfo {author} {\bibfnamefont {O.}~\bibnamefont {Vogt}},\
  }\href {\doibase 10.1016/0304-8853(85)90318-x} {\bibfield  {journal}
  {\bibinfo  {journal} {J. Magn. Magn. Mater.}\ }\textbf {\bibinfo {volume}
  {52}},\ \bibinfo {pages} {410} (\bibinfo {year} {1985})}\BibitemShut
  {NoStop}%
\bibitem [{\citenamefont {Osborn}\ \emph {et~al.}(1988)\citenamefont {Osborn},
  \citenamefont {Hagen}, \citenamefont {Jones}, \citenamefont {Stirling},
  \citenamefont {Lander}, \citenamefont {Mattenberger},\ and\ \citenamefont
  {Vogt}}]{Osborn1988High}%
  \BibitemOpen
  \bibfield  {author} {\bibinfo {author} {\bibfnamefont {R.}~\bibnamefont
  {Osborn}}, \bibinfo {author} {\bibfnamefont {M.}~\bibnamefont {Hagen}},
  \bibinfo {author} {\bibfnamefont {D.~L.}\ \bibnamefont {Jones}}, \bibinfo
  {author} {\bibfnamefont {W.~G.}\ \bibnamefont {Stirling}}, \bibinfo {author}
  {\bibfnamefont {G.~H.}\ \bibnamefont {Lander}}, \bibinfo {author}
  {\bibfnamefont {K.}~\bibnamefont {Mattenberger}}, \ and\ \bibinfo {author}
  {\bibfnamefont {O.}~\bibnamefont {Vogt}},\ }\href {\doibase
  10.1016/0304-8853(88)90445-3} {\bibfield  {journal} {\bibinfo  {journal} {J.
  Magn. Magn. Mater.}\ }\textbf {\bibinfo {volume} {76}},\ \bibinfo {pages}
  {429} (\bibinfo {year} {1988})}\BibitemShut {NoStop}%
\bibitem [{\citenamefont {Braithwaite}\ \emph {et~al.}(1996)\citenamefont
  {Braithwaite}, \citenamefont {Goncharenko}, \citenamefont {Mignot},
  \citenamefont {Ochiai},\ and\ \citenamefont {Vogt}}]{Magnetic1996pressure}%
  \BibitemOpen
  \bibfield  {author} {\bibinfo {author} {\bibfnamefont {D.}~\bibnamefont
  {Braithwaite}}, \bibinfo {author} {\bibfnamefont {I.~N.}\ \bibnamefont
  {Goncharenko}}, \bibinfo {author} {\bibfnamefont {J.~M.}\ \bibnamefont
  {Mignot}}, \bibinfo {author} {\bibfnamefont {A.}~\bibnamefont {Ochiai}}, \
  and\ \bibinfo {author} {\bibfnamefont {O.}~\bibnamefont {Vogt}},\ }\href
  {\doibase 10.1209/epl/i1996-00542-51} {\bibfield  {journal} {\bibinfo
  {journal} {Europhys. Lett.}\ }\textbf {\bibinfo {volume} {35}},\ \bibinfo
  {pages} {121} (\bibinfo {year} {1996})}\BibitemShut {NoStop}%
\bibitem [{\citenamefont {Braithwaite}\ and\ \citenamefont
  {Demuer}(1998)}]{Braithwaite1998Pressure}%
  \BibitemOpen
  \bibfield  {author} {\bibinfo {author} {\bibfnamefont {D.}~\bibnamefont
  {Braithwaite}}\ and\ \bibinfo {author} {\bibfnamefont {A.}~\bibnamefont
  {Demuer}},\ }\href {\doibase 10.1016/s0925-8388(98)00104-2} {\bibfield
  {journal} {\bibinfo  {journal} {J. Alloys Compd.}\ }\textbf {\bibinfo
  {volume} {271}},\ \bibinfo {pages} {426} (\bibinfo {year}
  {1998})}\BibitemShut {NoStop}%
\bibitem [{\citenamefont {Nuttall}\ \emph {et~al.}(2002)\citenamefont
  {Nuttall}, \citenamefont {Perry}, \citenamefont {Stirling}, \citenamefont
  {Mitchell}, \citenamefont {Kilcoyne},\ and\ \citenamefont
  {Cywinski}}]{Nuttall2002mag}%
  \BibitemOpen
  \bibfield  {author} {\bibinfo {author} {\bibfnamefont {W.~J.}\ \bibnamefont
  {Nuttall}}, \bibinfo {author} {\bibfnamefont {S.~C.}\ \bibnamefont {Perry}},
  \bibinfo {author} {\bibfnamefont {W.~G.}\ \bibnamefont {Stirling}}, \bibinfo
  {author} {\bibfnamefont {P.~W.}\ \bibnamefont {Mitchell}}, \bibinfo {author}
  {\bibfnamefont {S.~H.}\ \bibnamefont {Kilcoyne}}, \ and\ \bibinfo {author}
  {\bibfnamefont {R.}~\bibnamefont {Cywinski}},\ }\href {\doibase
  10.1016/s0921-4526(01)01034-1} {\bibfield  {journal} {\bibinfo  {journal}
  {Phys. B Condens. Matter}\ }\textbf {\bibinfo {volume} {315}},\ \bibinfo
  {pages} {179} (\bibinfo {year} {2002})}\BibitemShut {NoStop}%
\bibitem [{\citenamefont {Lim}\ \emph {et~al.}(2013)\citenamefont {Lim},
  \citenamefont {Blackburn}, \citenamefont {Magnani}, \citenamefont {Hiess},
  \citenamefont {Regnault}, \citenamefont {Caciuffo},\ and\ \citenamefont
  {Lander}}]{PhysRevB.87.064421}%
  \BibitemOpen
  \bibfield  {author} {\bibinfo {author} {\bibfnamefont {J.~A.}\ \bibnamefont
  {Lim}}, \bibinfo {author} {\bibfnamefont {E.}~\bibnamefont {Blackburn}},
  \bibinfo {author} {\bibfnamefont {N.}~\bibnamefont {Magnani}}, \bibinfo
  {author} {\bibfnamefont {A.}~\bibnamefont {Hiess}}, \bibinfo {author}
  {\bibfnamefont {L.-P.}\ \bibnamefont {Regnault}}, \bibinfo {author}
  {\bibfnamefont {R.}~\bibnamefont {Caciuffo}}, \ and\ \bibinfo {author}
  {\bibfnamefont {G.~H.}\ \bibnamefont {Lander}},\ }\href {\doibase
  10.1103/PhysRevB.87.064421} {\bibfield  {journal} {\bibinfo  {journal} {Phys.
  Rev. B}\ }\textbf {\bibinfo {volume} {87}},\ \bibinfo {pages} {064421}
  (\bibinfo {year} {2013})}\BibitemShut {NoStop}%
\bibitem [{\citenamefont {Monachesi}\ and\ \citenamefont
  {Weling}(1983)}]{Monachesi1983Low}%
  \BibitemOpen
  \bibfield  {author} {\bibinfo {author} {\bibfnamefont {P.}~\bibnamefont
  {Monachesi}}\ and\ \bibinfo {author} {\bibfnamefont {F.}~\bibnamefont
  {Weling}},\ }\href {\doibase 10.1103/PhysRevB.28.270} {\bibfield  {journal}
  {\bibinfo  {journal} {Phys. Rev. B}\ }\textbf {\bibinfo {volume} {28}},\
  \bibinfo {pages} {270} (\bibinfo {year} {1983})}\BibitemShut {NoStop}%
\bibitem [{\citenamefont {Kumigashira}\ \emph {et~al.}(1999)\citenamefont
  {Kumigashira}, \citenamefont {Kim}, \citenamefont {Ito}, \citenamefont
  {Ashihara}, \citenamefont {Takahashi}, \citenamefont {Aoki}, \citenamefont
  {Ochiai},\ and\ \citenamefont {Suzuki}}]{Kumigashira1999High}%
  \BibitemOpen
  \bibfield  {author} {\bibinfo {author} {\bibfnamefont {H.}~\bibnamefont
  {Kumigashira}}, \bibinfo {author} {\bibfnamefont {H.~D.}\ \bibnamefont
  {Kim}}, \bibinfo {author} {\bibfnamefont {T.}~\bibnamefont {Ito}}, \bibinfo
  {author} {\bibfnamefont {A.}~\bibnamefont {Ashihara}}, \bibinfo {author}
  {\bibfnamefont {T.}~\bibnamefont {Takahashi}}, \bibinfo {author}
  {\bibfnamefont {H.}~\bibnamefont {Aoki}}, \bibinfo {author} {\bibfnamefont
  {A.}~\bibnamefont {Ochiai}}, \ and\ \bibinfo {author} {\bibfnamefont
  {T.}~\bibnamefont {Suzuki}},\ }\href {\doibase 10.1016/S0921-4526(98)01033-3}
  {\bibfield  {journal} {\bibinfo  {journal} {Phys. B Condens. Matter}\
  }\textbf {\bibinfo {volume} {259}},\ \bibinfo {pages} {1124} (\bibinfo {year}
  {1999})}\BibitemShut {NoStop}%
\bibitem [{\citenamefont {Kumigashira}\ \emph
  {et~al.}(2000{\natexlab{a}})\citenamefont {Kumigashira}, \citenamefont {Ito},
  \citenamefont {Ashihara}, \citenamefont {Kim}, \citenamefont {Aoki},
  \citenamefont {Suzuki}, \citenamefont {Yamagami}, \citenamefont {Takahashi},\
  and\ \citenamefont {Ochiai}}]{PhysRevB.61.15707}%
  \BibitemOpen
  \bibfield  {author} {\bibinfo {author} {\bibfnamefont {H.}~\bibnamefont
  {Kumigashira}}, \bibinfo {author} {\bibfnamefont {T.}~\bibnamefont {Ito}},
  \bibinfo {author} {\bibfnamefont {A.}~\bibnamefont {Ashihara}}, \bibinfo
  {author} {\bibfnamefont {H.-D.}\ \bibnamefont {Kim}}, \bibinfo {author}
  {\bibfnamefont {H.}~\bibnamefont {Aoki}}, \bibinfo {author} {\bibfnamefont
  {T.}~\bibnamefont {Suzuki}}, \bibinfo {author} {\bibfnamefont
  {H.}~\bibnamefont {Yamagami}}, \bibinfo {author} {\bibfnamefont
  {T.}~\bibnamefont {Takahashi}}, \ and\ \bibinfo {author} {\bibfnamefont
  {A.}~\bibnamefont {Ochiai}},\ }\href {\doibase 10.1103/PhysRevB.61.15707}
  {\bibfield  {journal} {\bibinfo  {journal} {Phys. Rev. B}\ }\textbf {\bibinfo
  {volume} {61}},\ \bibinfo {pages} {15707} (\bibinfo {year}
  {2000}{\natexlab{a}})}\BibitemShut {NoStop}%
\bibitem [{\citenamefont {Yamagami}(2000)}]{PhysRevB.61.6246}%
  \BibitemOpen
  \bibfield  {author} {\bibinfo {author} {\bibfnamefont {H.}~\bibnamefont
  {Yamagami}},\ }\href {\doibase 10.1103/PhysRevB.61.6246} {\bibfield
  {journal} {\bibinfo  {journal} {Phys. Rev. B}\ }\textbf {\bibinfo {volume}
  {61}},\ \bibinfo {pages} {6246} (\bibinfo {year} {2000})}\BibitemShut
  {NoStop}%
\bibitem [{\citenamefont {Baptist}\ \emph {et~al.}(1980)\citenamefont
  {Baptist}, \citenamefont {Belakhovsky}, \citenamefont {Brooks}, \citenamefont
  {Pinchaux}, \citenamefont {Baer},\ and\ \citenamefont
  {Vogt}}]{BAPTIST198063}%
  \BibitemOpen
  \bibfield  {author} {\bibinfo {author} {\bibfnamefont {R.}~\bibnamefont
  {Baptist}}, \bibinfo {author} {\bibfnamefont {M.}~\bibnamefont
  {Belakhovsky}}, \bibinfo {author} {\bibfnamefont {M.~S.~S.}\ \bibnamefont
  {Brooks}}, \bibinfo {author} {\bibfnamefont {R.}~\bibnamefont {Pinchaux}},
  \bibinfo {author} {\bibfnamefont {Y.}~\bibnamefont {Baer}}, \ and\ \bibinfo
  {author} {\bibfnamefont {O.}~\bibnamefont {Vogt}},\ }\href {\doibase
  https://doi.org/10.1016/0378-4363(80)90128-X} {\bibfield  {journal} {\bibinfo
   {journal} {Physica B+C}\ }\textbf {\bibinfo {volume} {102}},\ \bibinfo
  {pages} {63 } (\bibinfo {year} {1980})}\BibitemShut {NoStop}%
\bibitem [{\citenamefont {Erbudak}\ and\ \citenamefont
  {Meier}(1980)}]{ERBUDAK1980134}%
  \BibitemOpen
  \bibfield  {author} {\bibinfo {author} {\bibfnamefont {M.}~\bibnamefont
  {Erbudak}}\ and\ \bibinfo {author} {\bibfnamefont {F.}~\bibnamefont
  {Meier}},\ }\href {\doibase https://doi.org/10.1016/0378-4363(80)90143-6}
  {\bibfield  {journal} {\bibinfo  {journal} {Physica B+C}\ }\textbf {\bibinfo
  {volume} {102}},\ \bibinfo {pages} {134 } (\bibinfo {year}
  {1980})}\BibitemShut {NoStop}%
\bibitem [{\citenamefont {Tang}\ \emph {et~al.}(1992)\citenamefont {Tang},
  \citenamefont {Stirling}, \citenamefont {Lander}, \citenamefont {Gibbs},
  \citenamefont {Herzog}, \citenamefont {Carra}, \citenamefont {Thole},
  \citenamefont {Mattenberger},\ and\ \citenamefont {Vogt}}]{Tang1992Resonant}%
  \BibitemOpen
  \bibfield  {author} {\bibinfo {author} {\bibfnamefont {C.~C.}\ \bibnamefont
  {Tang}}, \bibinfo {author} {\bibfnamefont {W.~G.}\ \bibnamefont {Stirling}},
  \bibinfo {author} {\bibfnamefont {G.~H.}\ \bibnamefont {Lander}}, \bibinfo
  {author} {\bibfnamefont {D.}~\bibnamefont {Gibbs}}, \bibinfo {author}
  {\bibfnamefont {W.}~\bibnamefont {Herzog}}, \bibinfo {author} {\bibfnamefont
  {P.}~\bibnamefont {Carra}}, \bibinfo {author} {\bibfnamefont {B.~T.}\
  \bibnamefont {Thole}}, \bibinfo {author} {\bibfnamefont {K.}~\bibnamefont
  {Mattenberger}}, \ and\ \bibinfo {author} {\bibfnamefont {O.}~\bibnamefont
  {Vogt}},\ }\href {\doibase 10.1103/PhysRevB.46.5287} {\bibfield  {journal}
  {\bibinfo  {journal} {Phys. Rev. B}\ }\textbf {\bibinfo {volume} {46}},\
  \bibinfo {pages} {5287} (\bibinfo {year} {1992})}\BibitemShut {NoStop}%
\bibitem [{\citenamefont {Kumigashira}\ \emph
  {et~al.}(2000{\natexlab{b}})\citenamefont {Kumigashira}, \citenamefont {Ito},
  \citenamefont {Ashihara}, \citenamefont {Kim}, \citenamefont {Aoki},
  \citenamefont {Suzuki}, \citenamefont {Yamagami}, \citenamefont {Takahashi},\
  and\ \citenamefont {Ochiai}}]{Kumigashira2000High}%
  \BibitemOpen
  \bibfield  {author} {\bibinfo {author} {\bibfnamefont {H.}~\bibnamefont
  {Kumigashira}}, \bibinfo {author} {\bibfnamefont {T.}~\bibnamefont {Ito}},
  \bibinfo {author} {\bibfnamefont {A.}~\bibnamefont {Ashihara}}, \bibinfo
  {author} {\bibfnamefont {H.~D.}\ \bibnamefont {Kim}}, \bibinfo {author}
  {\bibfnamefont {H.}~\bibnamefont {Aoki}}, \bibinfo {author} {\bibfnamefont
  {T.}~\bibnamefont {Suzuki}}, \bibinfo {author} {\bibfnamefont
  {H.}~\bibnamefont {Yamagami}}, \bibinfo {author} {\bibfnamefont
  {T.}~\bibnamefont {Takahashi}}, \ and\ \bibinfo {author} {\bibfnamefont
  {A.}~\bibnamefont {Ochiai}},\ }\href {\doibase 10.1016/s0304-8853(00)00631-4}
  {\bibfield  {journal} {\bibinfo  {journal} {J. Magn. Magn. Mater.}\ }\textbf
  {\bibinfo {volume} {226}},\ \bibinfo {pages} {68} (\bibinfo {year}
  {2000}{\natexlab{b}})}\BibitemShut {NoStop}%
\bibitem [{\citenamefont {Suzuki}\ \emph {et~al.}(1995)\citenamefont {Suzuki},
  \citenamefont {Hotta}, \citenamefont {Haga}, \citenamefont {Ochiai},
  \citenamefont {Suzuki}, \citenamefont {Shikama},\ and\ \citenamefont
  {Suzuki}}]{Suzuki1995}%
  \BibitemOpen
  \bibfield  {author} {\bibinfo {author} {\bibfnamefont {T.}~\bibnamefont
  {Suzuki}}, \bibinfo {author} {\bibfnamefont {E.}~\bibnamefont {Hotta}},
  \bibinfo {author} {\bibfnamefont {Y.}~\bibnamefont {Haga}}, \bibinfo {author}
  {\bibfnamefont {A.}~\bibnamefont {Ochiai}}, \bibinfo {author} {\bibfnamefont
  {Y.}~\bibnamefont {Suzuki}}, \bibinfo {author} {\bibfnamefont
  {T.}~\bibnamefont {Shikama}}, \ and\ \bibinfo {author} {\bibfnamefont
  {K.}~\bibnamefont {Suzuki}},\ }\href {\doibase 10.1016/0925-8388(94)05002-3}
  {\bibfield  {journal} {\bibinfo  {journal} {J. Alloys Compd.}\ }\textbf
  {\bibinfo {volume} {219}},\ \bibinfo {pages} {252–255} (\bibinfo {year}
  {1995})}\BibitemShut {NoStop}%
\bibitem [{\citenamefont {Ishiguro}\ \emph {et~al.}(1997)\citenamefont
  {Ishiguro}, \citenamefont {Aoki}, \citenamefont {Sugie}, \citenamefont
  {Suzuki}, \citenamefont {Sawada}, \citenamefont {Sato}, \citenamefont
  {Komatsubara}, \citenamefont {Ochiai}, \citenamefont {Suzuki}, \citenamefont
  {Suzuki}, \citenamefont {Higuchi},\ and\ \citenamefont
  {Hasegawa}}]{JPSJ.66.2764}%
  \BibitemOpen
  \bibfield  {author} {\bibinfo {author} {\bibfnamefont {A.}~\bibnamefont
  {Ishiguro}}, \bibinfo {author} {\bibfnamefont {H.}~\bibnamefont {Aoki}},
  \bibinfo {author} {\bibfnamefont {O.}~\bibnamefont {Sugie}}, \bibinfo
  {author} {\bibfnamefont {M.}~\bibnamefont {Suzuki}}, \bibinfo {author}
  {\bibfnamefont {A.}~\bibnamefont {Sawada}}, \bibinfo {author} {\bibfnamefont
  {N.}~\bibnamefont {Sato}}, \bibinfo {author} {\bibfnamefont {T.}~\bibnamefont
  {Komatsubara}}, \bibinfo {author} {\bibfnamefont {A.}~\bibnamefont {Ochiai}},
  \bibinfo {author} {\bibfnamefont {T.}~\bibnamefont {Suzuki}}, \bibinfo
  {author} {\bibfnamefont {K.}~\bibnamefont {Suzuki}}, \bibinfo {author}
  {\bibfnamefont {M.}~\bibnamefont {Higuchi}}, \ and\ \bibinfo {author}
  {\bibfnamefont {A.}~\bibnamefont {Hasegawa}},\ }\href {\doibase
  10.1143/JPSJ.66.2764} {\bibfield  {journal} {\bibinfo  {journal} {J. Phys.
  Soc. Jpn.}\ }\textbf {\bibinfo {volume} {66}},\ \bibinfo {pages} {2764}
  (\bibinfo {year} {1997})}\BibitemShut {NoStop}%
\bibitem [{\citenamefont {Schoenes}(1981)}]{Schoenes1981}%
  \BibitemOpen
  \bibfield  {author} {\bibinfo {author} {\bibfnamefont {J.}~\bibnamefont
  {Schoenes}},\ }\href {\doibase 10.1016/0370-1573(80)90156-8} {\bibfield
  {journal} {\bibinfo  {journal} {Phys. Rep.}\ }\textbf {\bibinfo {volume}
  {66}},\ \bibinfo {pages} {187} (\bibinfo {year} {1981})}\BibitemShut
  {NoStop}%
\bibitem [{\citenamefont {Weinberger}\ and\ \citenamefont
  {Podloucky}(1980)}]{PhysRevB.22.645}%
  \BibitemOpen
  \bibfield  {author} {\bibinfo {author} {\bibfnamefont {P.}~\bibnamefont
  {Weinberger}}\ and\ \bibinfo {author} {\bibfnamefont {R.}~\bibnamefont
  {Podloucky}},\ }\href {\doibase 10.1103/PhysRevB.22.645} {\bibfield
  {journal} {\bibinfo  {journal} {Phys. Rev. B}\ }\textbf {\bibinfo {volume}
  {22}},\ \bibinfo {pages} {645} (\bibinfo {year} {1980})}\BibitemShut
  {NoStop}%
\bibitem [{\citenamefont {Kasuya}(1995)}]{JPSJ.64.2294}%
  \BibitemOpen
  \bibfield  {author} {\bibinfo {author} {\bibfnamefont {T.}~\bibnamefont
  {Kasuya}},\ }\href {\doibase 10.1143/JPSJ.64.2294} {\bibfield  {journal}
  {\bibinfo  {journal} {J. Phys. Soc. Jpn.}\ }\textbf {\bibinfo {volume}
  {64}},\ \bibinfo {pages} {2294} (\bibinfo {year} {1995})}\BibitemShut
  {NoStop}%
\bibitem [{\citenamefont {Kasuya}(1996)}]{JPSJ.65.3394}%
  \BibitemOpen
  \bibfield  {author} {\bibinfo {author} {\bibfnamefont {T.}~\bibnamefont
  {Kasuya}},\ }\href {\doibase 10.1143/JPSJ.65.3394} {\bibfield  {journal}
  {\bibinfo  {journal} {J. Phys. Soc. Jpn.}\ }\textbf {\bibinfo {volume}
  {65}},\ \bibinfo {pages} {3394} (\bibinfo {year} {1996})}\BibitemShut
  {NoStop}%
\bibitem [{\citenamefont {Georges}\ \emph {et~al.}(1996)\citenamefont
  {Georges}, \citenamefont {Kotliar}, \citenamefont {Krauth},\ and\
  \citenamefont {Rozenberg}}]{RevModPhys.68.13}%
  \BibitemOpen
  \bibfield  {author} {\bibinfo {author} {\bibfnamefont {A.}~\bibnamefont
  {Georges}}, \bibinfo {author} {\bibfnamefont {G.}~\bibnamefont {Kotliar}},
  \bibinfo {author} {\bibfnamefont {W.}~\bibnamefont {Krauth}}, \ and\ \bibinfo
  {author} {\bibfnamefont {M.~J.}\ \bibnamefont {Rozenberg}},\ }\href {\doibase
  10.1103/RevModPhys.68.13} {\bibfield  {journal} {\bibinfo  {journal} {Rev.
  Mod. Phys.}\ }\textbf {\bibinfo {volume} {68}},\ \bibinfo {pages} {13}
  (\bibinfo {year} {1996})}\BibitemShut {NoStop}%
\bibitem [{\citenamefont {Goremychkin}\ \emph {et~al.}(2018)\citenamefont
  {Goremychkin}, \citenamefont {Park}, \citenamefont {Osborn}, \citenamefont
  {Rosenkranz}, \citenamefont {Castellan}, \citenamefont {Fanelli},
  \citenamefont {Christianson}, \citenamefont {Stone}, \citenamefont {Bauer},
  \citenamefont {McClellan}, \citenamefont {Byler},\ and\ \citenamefont
  {Lawrence}}]{Goremychkin186}%
  \BibitemOpen
  \bibfield  {author} {\bibinfo {author} {\bibfnamefont {E.~A.}\ \bibnamefont
  {Goremychkin}}, \bibinfo {author} {\bibfnamefont {H.}~\bibnamefont {Park}},
  \bibinfo {author} {\bibfnamefont {R.}~\bibnamefont {Osborn}}, \bibinfo
  {author} {\bibfnamefont {S.}~\bibnamefont {Rosenkranz}}, \bibinfo {author}
  {\bibfnamefont {J.-P.}\ \bibnamefont {Castellan}}, \bibinfo {author}
  {\bibfnamefont {V.~R.}\ \bibnamefont {Fanelli}}, \bibinfo {author}
  {\bibfnamefont {A.~D.}\ \bibnamefont {Christianson}}, \bibinfo {author}
  {\bibfnamefont {M.~B.}\ \bibnamefont {Stone}}, \bibinfo {author}
  {\bibfnamefont {E.~D.}\ \bibnamefont {Bauer}}, \bibinfo {author}
  {\bibfnamefont {K.~J.}\ \bibnamefont {McClellan}}, \bibinfo {author}
  {\bibfnamefont {D.~D.}\ \bibnamefont {Byler}}, \ and\ \bibinfo {author}
  {\bibfnamefont {J.~M.}\ \bibnamefont {Lawrence}},\ }\href {\doibase
  10.1126/science.aan0593} {\bibfield  {journal} {\bibinfo  {journal}
  {Science}\ }\textbf {\bibinfo {volume} {359}},\ \bibinfo {pages} {186}
  (\bibinfo {year} {2018})}\BibitemShut {NoStop}%
\bibitem [{\citenamefont {Haule}\ and\ \citenamefont
  {Kotliar}(2009)}]{Haule2009Arrested}%
  \BibitemOpen
  \bibfield  {author} {\bibinfo {author} {\bibfnamefont {K.}~\bibnamefont
  {Haule}}\ and\ \bibinfo {author} {\bibfnamefont {G.}~\bibnamefont
  {Kotliar}},\ }\href {\doibase 10.1038/nphys1392} {\bibfield  {journal}
  {\bibinfo  {journal} {Nat. Phys.}\ }\textbf {\bibinfo {volume} {5}},\
  \bibinfo {pages} {796} (\bibinfo {year} {2009})}\BibitemShut {NoStop}%
\bibitem [{\citenamefont {Blaha}\ \emph {et~al.}(2001)\citenamefont {Blaha},
  \citenamefont {Schwarz}, \citenamefont {Madsen}, \citenamefont {Kvasnicka},\
  and\ \citenamefont {Luitz}}]{wien2k}%
  \BibitemOpen
  \bibfield  {author} {\bibinfo {author} {\bibfnamefont {P.}~\bibnamefont
  {Blaha}}, \bibinfo {author} {\bibfnamefont {K.}~\bibnamefont {Schwarz}},
  \bibinfo {author} {\bibfnamefont {G.}~\bibnamefont {Madsen}}, \bibinfo
  {author} {\bibfnamefont {D.}~\bibnamefont {Kvasnicka}}, \ and\ \bibinfo
  {author} {\bibfnamefont {J.}~\bibnamefont {Luitz}},\ }\href@noop {} {\emph
  {\bibinfo {title} {{WIEN2k, An Augmented Plane Wave + Local Orbitals Program
  for Calculating Crystal Properties}}}}\ (\bibinfo  {publisher} {Karlheinz
  Schwarz, Techn. Universität Wien, Austria},\ \bibinfo {year}
  {2001})\BibitemShut {NoStop}%
\bibitem [{\citenamefont {Perdew}\ \emph {et~al.}(1996)\citenamefont {Perdew},
  \citenamefont {Burke},\ and\ \citenamefont
  {Ernzerhof}}]{PhysRevLett.77.3865}%
  \BibitemOpen
  \bibfield  {author} {\bibinfo {author} {\bibfnamefont {J.~P.}\ \bibnamefont
  {Perdew}}, \bibinfo {author} {\bibfnamefont {K.}~\bibnamefont {Burke}}, \
  and\ \bibinfo {author} {\bibfnamefont {M.}~\bibnamefont {Ernzerhof}},\ }\href
  {\doibase 10.1103/PhysRevLett.77.3865} {\bibfield  {journal} {\bibinfo
  {journal} {Phys. Rev. Lett.}\ }\textbf {\bibinfo {volume} {77}},\ \bibinfo
  {pages} {3865} (\bibinfo {year} {1996})}\BibitemShut {NoStop}%
\bibitem [{\citenamefont {Haule}\ \emph {et~al.}(2010)\citenamefont {Haule},
  \citenamefont {Yee},\ and\ \citenamefont {Kim}}]{PhysRevB.81.195107}%
  \BibitemOpen
  \bibfield  {author} {\bibinfo {author} {\bibfnamefont {K.}~\bibnamefont
  {Haule}}, \bibinfo {author} {\bibfnamefont {C.-H.}\ \bibnamefont {Yee}}, \
  and\ \bibinfo {author} {\bibfnamefont {K.}~\bibnamefont {Kim}},\ }\href
  {\doibase 10.1103/PhysRevB.81.195107} {\bibfield  {journal} {\bibinfo
  {journal} {Phys. Rev. B}\ }\textbf {\bibinfo {volume} {81}},\ \bibinfo
  {pages} {195107} (\bibinfo {year} {2010})}\BibitemShut {NoStop}%
\bibitem [{\citenamefont {Gull}\ \emph {et~al.}(2011)\citenamefont {Gull},
  \citenamefont {Millis}, \citenamefont {Lichtenstein}, \citenamefont
  {Rubtsov}, \citenamefont {Troyer},\ and\ \citenamefont
  {Werner}}]{RevModPhys.83.349}%
  \BibitemOpen
  \bibfield  {author} {\bibinfo {author} {\bibfnamefont {E.}~\bibnamefont
  {Gull}}, \bibinfo {author} {\bibfnamefont {A.~J.}\ \bibnamefont {Millis}},
  \bibinfo {author} {\bibfnamefont {A.~I.}\ \bibnamefont {Lichtenstein}},
  \bibinfo {author} {\bibfnamefont {A.~N.}\ \bibnamefont {Rubtsov}}, \bibinfo
  {author} {\bibfnamefont {M.}~\bibnamefont {Troyer}}, \ and\ \bibinfo {author}
  {\bibfnamefont {P.}~\bibnamefont {Werner}},\ }\href {\doibase
  10.1103/RevModPhys.83.349} {\bibfield  {journal} {\bibinfo  {journal} {Rev.
  Mod. Phys.}\ }\textbf {\bibinfo {volume} {83}},\ \bibinfo {pages} {349}
  (\bibinfo {year} {2011})}\BibitemShut {NoStop}%
\bibitem [{\citenamefont {Werner}\ \emph {et~al.}(2006)\citenamefont {Werner},
  \citenamefont {Comanac}, \citenamefont {de' Medici}, \citenamefont {Troyer},\
  and\ \citenamefont {Millis}}]{PhysRevLett.97.076405}%
  \BibitemOpen
  \bibfield  {author} {\bibinfo {author} {\bibfnamefont {P.}~\bibnamefont
  {Werner}}, \bibinfo {author} {\bibfnamefont {A.}~\bibnamefont {Comanac}},
  \bibinfo {author} {\bibfnamefont {L.}~\bibnamefont {de' Medici}}, \bibinfo
  {author} {\bibfnamefont {M.}~\bibnamefont {Troyer}}, \ and\ \bibinfo {author}
  {\bibfnamefont {A.~J.}\ \bibnamefont {Millis}},\ }\href {\doibase
  10.1103/PhysRevLett.97.076405} {\bibfield  {journal} {\bibinfo  {journal}
  {Phys. Rev. Lett.}\ }\textbf {\bibinfo {volume} {97}},\ \bibinfo {pages}
  {076405} (\bibinfo {year} {2006})}\BibitemShut {NoStop}%
\bibitem [{\citenamefont {Baer}\ \emph {et~al.}(1978)\citenamefont {Baer},
  \citenamefont {Hauger}, \citenamefont {Z\"urcher}, \citenamefont {Campagna},\
  and\ \citenamefont {Wertheim}}]{PhysRevB.18.4433}%
  \BibitemOpen
  \bibfield  {author} {\bibinfo {author} {\bibfnamefont {Y.}~\bibnamefont
  {Baer}}, \bibinfo {author} {\bibfnamefont {R.}~\bibnamefont {Hauger}},
  \bibinfo {author} {\bibfnamefont {C.}~\bibnamefont {Z\"urcher}}, \bibinfo
  {author} {\bibfnamefont {M.}~\bibnamefont {Campagna}}, \ and\ \bibinfo
  {author} {\bibfnamefont {G.~K.}\ \bibnamefont {Wertheim}},\ }\href {\doibase
  10.1103/PhysRevB.18.4433} {\bibfield  {journal} {\bibinfo  {journal} {Phys.
  Rev. B}\ }\textbf {\bibinfo {volume} {18}},\ \bibinfo {pages} {4433}
  (\bibinfo {year} {1978})}\BibitemShut {NoStop}%
\bibitem [{\citenamefont {Reihl}\ \emph {et~al.}(1982)\citenamefont {Reihl},
  \citenamefont {Martensson},\ and\ \citenamefont
  {Vogt}}]{Reihl1982Electronic}%
  \BibitemOpen
  \bibfield  {author} {\bibinfo {author} {\bibfnamefont {B.}~\bibnamefont
  {Reihl}}, \bibinfo {author} {\bibfnamefont {N.}~\bibnamefont {Martensson}}, \
  and\ \bibinfo {author} {\bibfnamefont {O.}~\bibnamefont {Vogt}},\ }\href
  {\doibase 10.1063/1.330689} {\bibfield  {journal} {\bibinfo  {journal} {J.
  Appl. Phys.}\ }\textbf {\bibinfo {volume} {53}},\ \bibinfo {pages} {2008}
  (\bibinfo {year} {1982})}\BibitemShut {NoStop}%
\bibitem [{\citenamefont {Reihl}(1987)}]{Reihl1987Photoelectron}%
  \BibitemOpen
  \bibfield  {author} {\bibinfo {author} {\bibfnamefont {B.}~\bibnamefont
  {Reihl}},\ }\href {\doibase 10.1016/0022-5088(87)90220-7} {\bibfield
  {journal} {\bibinfo  {journal} {J. Less-Common Met.}\ }\textbf {\bibinfo
  {volume} {128}},\ \bibinfo {pages} {331} (\bibinfo {year}
  {1987})}\BibitemShut {NoStop}%
\bibitem [{\citenamefont {Choi}\ \emph {et~al.}(2012)\citenamefont {Choi},
  \citenamefont {Min}, \citenamefont {Shim}, \citenamefont {Haule},\ and\
  \citenamefont {Kotliar}}]{PhysRevLett.108.016402}%
  \BibitemOpen
  \bibfield  {author} {\bibinfo {author} {\bibfnamefont {H.~C.}\ \bibnamefont
  {Choi}}, \bibinfo {author} {\bibfnamefont {B.~I.}\ \bibnamefont {Min}},
  \bibinfo {author} {\bibfnamefont {J.~H.}\ \bibnamefont {Shim}}, \bibinfo
  {author} {\bibfnamefont {K.}~\bibnamefont {Haule}}, \ and\ \bibinfo {author}
  {\bibfnamefont {G.}~\bibnamefont {Kotliar}},\ }\href {\doibase
  10.1103/PhysRevLett.108.016402} {\bibfield  {journal} {\bibinfo  {journal}
  {Phys. Rev. Lett.}\ }\textbf {\bibinfo {volume} {108}},\ \bibinfo {pages}
  {016402} (\bibinfo {year} {2012})}\BibitemShut {NoStop}%
\bibitem [{\citenamefont {Shim}\ \emph {et~al.}(2007)\citenamefont {Shim},
  \citenamefont {Haule},\ and\ \citenamefont {Kotliar}}]{shim:1615}%
  \BibitemOpen
  \bibfield  {author} {\bibinfo {author} {\bibfnamefont {J.~H.}\ \bibnamefont
  {Shim}}, \bibinfo {author} {\bibfnamefont {K.}~\bibnamefont {Haule}}, \ and\
  \bibinfo {author} {\bibfnamefont {G.}~\bibnamefont {Kotliar}},\ }\href
  {\doibase 10.1126/science.1149064} {\bibfield  {journal} {\bibinfo  {journal}
  {Science}\ }\textbf {\bibinfo {volume} {318}},\ \bibinfo {pages} {1615}
  (\bibinfo {year} {2007})}\BibitemShut {NoStop}%
\bibitem [{\citenamefont {Lu}\ and\ \citenamefont
  {Huang}(2018)}]{PhysRevB.98.195102}%
  \BibitemOpen
  \bibfield  {author} {\bibinfo {author} {\bibfnamefont {H.}~\bibnamefont
  {Lu}}\ and\ \bibinfo {author} {\bibfnamefont {L.}~\bibnamefont {Huang}},\
  }\href {\doibase 10.1103/PhysRevB.98.195102} {\bibfield  {journal} {\bibinfo
  {journal} {Phys. Rev. B}\ }\textbf {\bibinfo {volume} {98}},\ \bibinfo
  {pages} {195102} (\bibinfo {year} {2018})}\BibitemShut {NoStop}%
\bibitem [{\citenamefont {Haule}(2007)}]{PhysRevB.75.155113}%
  \BibitemOpen
  \bibfield  {author} {\bibinfo {author} {\bibfnamefont {K.}~\bibnamefont
  {Haule}},\ }\href {\doibase 10.1103/PhysRevB.75.155113} {\bibfield  {journal}
  {\bibinfo  {journal} {Phys. Rev. B}\ }\textbf {\bibinfo {volume} {75}},\
  \bibinfo {pages} {155113} (\bibinfo {year} {2007})}\BibitemShut {NoStop}%
\end{thebibliography}%

\end{document}